\documentclass[a4paper,printer]{aa}

\usepackage{graphicx}
\usepackage{natbib}
\usepackage{txfonts}
\usepackage{color}
\usepackage{multirow}
\usepackage{stmaryrd}
\usepackage{siunitx}

\newcommand{\invers}[1]{{\sc invers{#1}}}

\newcommand{\llm}[1]{{\sc LLmodels}}

\newcommand{\bz}{$\langle B_\mathrm{z} \rangle$}
\newcommand{\kms}{km\,s$^{-1}$}

\newcommand{\abn}[2]{$\log (\mathrm{{#1}}/N_\mathrm{tot}) = $\,{#2}}

\newcommand{\cla}[1]{#1}
\newcommand{\clb}[1]{#1}

\graphicspath{{figures/}}

\begin{document}
\title{Magnetic Doppler \cla{imaging} of the chemically \\ peculiar star HD\,125248%
\thanks{Based on observations collected at the European Southern Observatory, Chile (ESO programs 088.D-0066, 090.D-0256).}}

\author{N.~Rusomarov\inst{1}
   \and O.~Kochukhov\inst{1}
   \and T.~Ryabchikova\inst{2}
   \and I.~Ilyin\inst{3}}

\institute{
Department of Physics and Astronomy, Uppsala University, Box 516, 75120 Uppsala, Sweden
\and
Institute of Astronomy, Russian Academy of Sciences, Pyatnistkaya 48, 119017 Moscow, Russia
\and
Astrophysikalisches Institut Potsdam, An der Sternwarte 16, 14482 Potsdam, Germany
}

\date{Received 9 November 2015, accepted 25 January 2016}

\titlerunning{Magnetic field and abundance spots of the Ap star HD\,125248}
\authorrunning{N.~Rusomarov et al.}

\abstract
{Intermediate-mass, chemically peculiar stars with strong magnetic fields give us an excellent opportunity to study the topology of their surface magnetic fields and the interplay between magnetic geometries and abundance inhomogeneities in the atmospheres of these stars.}
{We reconstruct detailed maps of the surface magnetic field and abundance distributions for the magnetic Ap star HD\,125248.}
{We performed the analysis based on phase-resolved, four Stokes parameter spectropolarimetric observations obtained with the HARPSpol instrument. These data were interpreted with the help of magnetic Doppler imaging technique and model atmospheres taking into account the effects of strong magnetic fields and non-solar chemical composition.}
{We improved the atmospheric parameters of the star, $T_\mathrm{eff} = 9850\pm250$\,K and \cla{$\log g = 4.05\pm0.10$}. We performed detailed abundance analysis, which confirmed that HD\,125248 has abundances typical of other Ap stars, and discovered significant vertical stratification effects for the \ion{Fe}{ii} and \ion{Cr}{ii} ions. We computed LSD Stokes profiles using several line masks corresponding to Fe-peak and rare earth elements, and studied their behavior with rotational phase. Combining previous longitudinal field measurements with our own observations we improved the rotational period of the star $P_\mathrm{rot} = 9.29558 \pm 0.00006$\,d. Magnetic Doppler imaging of HD\,125248 showed that its magnetic field is mostly poloidal and quasi-dipolar with two large spots of different polarity and field strength. The chemical maps of Fe, Cr, Ce, Nd, Gd, and Ti show abundance contrasts of 0.9--3.5\,dex. Among these elements, the Fe abundance map does not show high-contrast features. Cr is overabundant around the negative magnetic pole and has 3.5\,dex abundance range. The rare earth elements and Ti are overabundant near the positive magnetic pole.}
{The magnetic field of HD\,125248 has strong deviations from the classical \cla{oblique} dipole field geometry. The comparison of the magnetic field topology of HD\,125248 with the results derived for other stars using four Stokes magnetic Doppler imaging suggests evidence that the field topology becomes simpler with increasing age. The abundance maps show weak correlation with the magnetic field geometry, but they do not agree with the theoretical atomic diffusion calculations, which predict element accumulation in the horizontal field regions.}


\maketitle

\section{Introduction}\label{sec:intro}
The star HD\,125248 (CS~Vir, HR\,5355) is one of the most outstanding objects in the class of magnetic chemically peculiar stars. This object was discovered by \citet{Morgan1931p24} to be a strong spectrum variable of the type A1p SrCrEu \citep{Renson2009p961} with a period of $\simeq$ 9.3~days \citep{Deutsch1947p283}. \citet{Morgan1931p24} observed that the absorption lines of \ion{Cr}{ii} and \ion{Eu}{ii} varied in anti-phase to each other, while lines of Fe and Ti showed hardly any variation. Observations by \citet{Babcock1947p260} showed that HD\,125248 possesses \cla{a} strong magnetic field, whose line-of-sight component varies with the rotational period of the star between approximately -2 and +2\,kG \citep{Babcock1951p1}. Since then a considerable amount of magnetic field measurements has been obtained for HD\,125248, spanning for more than half a century, the latest study being by \citet[][see Sect.~\ref{ssec:rot} for references]{Shorlin2002p637}.  An equally substantial number of photometric studies exists (see \citealt{Mikulavsek2007p10} for references), showing that HD125248 is variable with the same period in the visible and near infrared light, spectrum and magnetic field \citep[e.g.,][]{Catalano1992p203,Catalano1998p463,Leone2001p118}. Recent spectroscopic studies by \citet{Mathys1992p31} and \citet{Gonzalez1994p209} have also investigated the variation of the oxygen lines with the rotational period of the star.

HD\,125248 was the first star for which \citet{Stibbs1950p395} proposed the oblique rotator model (ORM) as a way to explain the observed variations of the photometric, spectral, and magnetic field observables. In this framework, the magnetic field is frozen into the atmosphere of the star, with its axis not aligned with the rotational axis of the star, resulting in a time-dependent magnetic configuration as seen by the observer. HD\,125248 is also the first object for which an attempt at spherical harmonics analysis was performed with the aim to produce surface maps of the abundance anomalies and magnetic field \citep{Deutsch1957p139}. The latest study of the magnetic field of HD\,125248 was produced by \citet[\citeyear{Bagnulo2002p1023}]{Bagnulo1999p865} using low-resolution circular polarization spectra (Stokes~$I$ and~$V$). The study found that the magnetic field of HD\,125248 shows strong deviations from \cla{an oblique dipole} field. However, \citet{Bagnulo1999p865} could not find a unique solution for the more complex model represented by a superposition of a dipole and a quadrupole.

It is evident that HD\,125248 is an excellent candidate for magnetic Doppler imaging \citep[MDI,][]{Piskunov2002p736,Kochukhov2002p868}. This method can simultaneously reconstruct the surface abundance distribution of different chemical elements and restore the vector magnetic field at the stellar surface. Magnetic Doppler imaging based on the full Stokes vector spectropolarimetric observations is particularly powerful as it does not require a priori information about the (global) geometry of the magnetic field, compared to MDI based only on circular polarization data. Recent four Stokes parameter MDI studies of magnetic Ap stars have revealed field topologies that have significant departures from \cla{oblique dipole models} and even contain some small-scale magnetic structures ($\alpha^2$~CVn, \citealt{Kochukhov10p13,Silvester2014p182}; HD\,32633, \citealt{Silvester2015p2163}; 53~Cam, \citealt{Kochukhov2004p613}).

We obtained high-resolution spectropolarimetric observations of HD\,125248 as part of our program aimed at observing Ap stars in all four Stokes parameters using the HARPSpol instrument \citep{Piskunov11p7} at the ESO 3.6-m telescope. The project will provide insight into the mechanism of atomic diffusion in the presence of magnetic fields \citep{Michaud1981p244,Alecian2010p53} that is believed to be responsible for the appearance of horizontal (spots) and vertical (stratification) chemical abundance inhomogeneities. Moreover, \citet{Braithwaite06p1077} have shown that stable magnetic fields can exist in the interiors of main-sequence stars with radiative envelopes. In this context, detailed empirical information on the geometry of magnetic fields of Ap/Bp stars gathered in a systematic approach is essential for understanding the underlying magnetohydrodynamic processes that lead to the generation and evolution of global magnetic fields in the interiors of main-sequence stars with radiative envelopes. 

The paper is organized as follows. Section~\ref{sec:obs} describes the observations. In Sect.~\ref{sec:fund} we derive the fundamental parameters of the star. Section~\ref{sec:mag} discusses the polarization signatures in the mean lines profiles, the measurements of the integral magnetic observables, and the rotational period search. Section~\ref{sec:mdi} introduces the principles of MDI, describes the choice of spectral lines used in the analysis, and the determination of rotational velocity and orientation of the rotational axis. The resulting magnetic and chemical maps are discussed in Sect.~\ref{sec:mdi-maps}. The summary and discussion are presented in Sect.~\ref{sec:sum}.

\section{Observations}\label{sec:obs}
Spectropolarimetric observations of HD\,125248 were obtained at the ESO \cla{3.6-m} telescope with the HARPS spectrograph \citep{Mayor2003p20} and its polarimetric unit HARPSpol \citep{Piskunov11p7,Snik2011p237}. We acquired 36 individual Stokes parameter observations spread over 12 observing nights for two consecutive years starting from 2012. These observing runs yielded \clb{a good} phase coverage over the entire rotational period of the star.

A spectropolarimetric observation of an individual Stokes parameter consists of four sub-exposures obtained with different orientations of the quarter-wave (Stokes $V$) or half-wave (Stokes $Q$ and $U$) retarder plate. From one such sequence of observations we derive an intensity spectrum (Stokes~$I$) and one more Stokes $Q$, $U$ or $V$ parameter, respectively. The eight extracted one-dimensional spectra corresponding to four sub-exposures were combined using the ratio method \citep{Donati97p658,Bagnulo2009p993}. This method has the advantage of automatically producing a null spectrum that can be used for assessing spurious polarization and crosstalks.

The spectra have resolving power, $\lambda / \Delta \lambda$, of approximately 110\,000 and a \cla{coverage of} 3780--6910\,\AA{} with an 80\,\AA{} gap located at 5290\,\AA{}. The exposure time of each sub-exposure during the first observing run in 2012 was set to 500\,s. In order to compensate for the worse seeing experienced during the second observing run in 2013 each sub-exposure had exposure time between 600 and 900 seconds. The signal-to-noise (S/N) ratio measured for all observations at 5500\,\AA{} is 150--350. We note that for two nights (22 and 23 Feb. 2013) the S/N ratio for the blue part of the spectrum is less than 100, which was caused by highly variable seeing. The journal of observations is provided in Table~\ref{tab:obsjournal}.
\begin{table*}
\caption{Journal of spectropolarimetric observations of HD\,125248.}
\centering
\begin{tabular}{lccccccccc}
\hline
\hline
\multirow{2}{*}{UT Date} & \multicolumn{3}{c}{HJD\,(2\,455\,000+)} & \multirow{2}{*}{$\overline{\varphi}$}& \multirow{2}{*}{$\delta\varphi$}& S/N & \multicolumn{3}{c}{Median S/N}\\
 & $Q$ & $U$ & $V$ &  &  & range & $Q$ & $U$ & $V$\\
\hline
28 Mar. 2012 & 1014.7988 & 1014.8235 & 1014.7735 & 0.704 & 0.003 & 136 -- 442 & 317 & 309 &326 \\
29 Mar. 2012 & 1015.7721 & 1015.7968 & 1015.7465 & 0.808 & 0.003 & 101 -- 402 & 277 & 304 &250 \\
30 Mar. 2012 & 1016.8004 & 1016.8251 & 1016.7749 & 0.919 & 0.003 & 122 -- 411 & 297 & 309 &292 \\
31 Mar. 2012 & 1017.7727 & 1017.7975 & 1017.7475 & 0.023 & 0.003 & 105 -- 389 & 264 & 294 &257 \\
01 Apr. 2012  & 1018.8119 & 1018.8555 & 1018.7772 & 0.136 & 0.006 & 100 -- 479 & 331 & 360 &228 \\
02 Apr. 2012  & 1019.7723 & 1019.8109 & 1019.7369 & 0.239 & 0.005 & 116 -- 479 & 358 & 297 &278 \\
22 Feb. 2013 & 1345.8471 & 1345.8774 & 1345.8161 & 0.317 & 0.004 &  91 -- 338 & 234 & 252 &225 \\
23 Feb. 2013 & 1346.8530 & 1346.8893 & 1346.8193 & 0.425 & 0.005 &  63 -- 314 & 155 & 209 &228 \\
24 Feb. 2013 & 1347.8547 & 1347.8933 & 1347.8174 & 0.533 & 0.005 & 104 -- 361 & 258 & 268 &258 \\
25 Feb. 2013 & 1348.8597 & 1348.9030 & 1348.8179 & 0.641 & 0.006 & 120 -- 449 & 307 & 331 &295 \\
26 Feb. 2013 & 1349.8598 & 1349.9007 & 1349.8213 & 0.749 & 0.005 & 116 -- 478 & 344 & 356 &288 \\
27 Feb. 2013 & 1350.8502 & 1350.8935 & 1350.8105 & 0.855 & 0.006 & 111 -- 398 & 290 & 296 &272 \\
\hline
\end{tabular}
\label{tab:obsjournal}
\tablefoot{First column gives the UT date at the beginning of each observing night. Heliocentric Julian Dates (HJD) at mid-exposure for each observed Stokes parameter are given in columns 2--4. Mean phase, $\overline{\varphi}$, and the maximum difference, $\delta\varphi$, between $\overline{\varphi}$ and phases of individual Stokes parameter observations are presented in columns 5--6. Rotational phases were calculated according to our improved ephemeris (Sect.~\ref{ssec:rot}). \clb{The range of the S/N for each group of Stokes parameter observations taken during one night is given in column 7. The median S/N for each individual Stokes parameter observation are presented in columns 8--10.} The median S/N was calculated using several \cla{echelle} orders around $\lambda=5500$\,\AA, where maximum counts were reached.}
\end{table*}

The observational and data processing techniques applied to the spectra for HD\,125248 are identical to the ones described by \citet{Rusomarov2013p8}. We refer the reader to that paper for detailed discussion of the observation procedures and reduction steps, such as extraction of the spectra, calculation of the Stokes parameters, normalization of the resulting spectra, as well as a more complete description of the instrument.

In addition to the HARPSpol four Stokes parameter spectra we have analyzed 9 circular polarization observations of HD\,125248 collected in June 2001 with the help of the SOFIN spectropolarimeter at the Nordic Optical Telescope. These data have resolving power of $\lambda / \Delta \lambda =70\,000$ and provide an incomplete wavelength coverage of the {4770--7090}~\AA{} interval. The typical peak S/N ratio of these spectra is {200--300} for $\lambda > 5000$~\AA{}. The reduction of the SOFIN spectra closely followed the procedure described by \citet{Lueftinger10p71}. Since the HARPSpol spectra already provided a good phase coverage with superior spectral resolution, the SOFIN data was used solely for determination of the mean longitudinal magnetic field.

\section{Fundamental parameters of HD\,125248}\label{sec:fund}
The most recent effective temperature estimate for HD\,125248, $T_\mathrm{eff} = 9500$\,K, was obtained by \citet{Lipski2008p481} from a fit of metal-enhanced model atmospheres to the observed spectral energy distributions (SED) in the ultraviolet (UV) and the optical regions. However, the authors assumed a surface gravity $\log g = 4.0$ and \cla{an} increased \cla{metallicity} by 0.5\,dex relative to the \cla{Sun}, which raises some concerns about the accuracy of the study as the phase averaged mean spectrum shows metal lines that are significantly stronger than what was assumed by \citet{Lipski2008p481}. In an earlier study, \citet{Monier1992p175} found that the energy distribution of HD\,125248  is best reproduced by an atmosphere model with a slightly higher temperature, $T_\mathrm{eff} = 9700$\,K, and a surface gravity $\log g = 4.25$ with $[M/H] = +1.0$\,dex. 

We started \cla{the} atmospheric analysis by computing a model atmosphere with the same values of $T_\mathrm{eff}$, $\log g$, and chemical composition as used by \citet{Lipski2008p481}, with the \llm{} code \citep{Shulyak2004p993}, which incorporates individual elemental abundances and has detailed treatment of line opacities due to Zeeman splitting and polarized radiative transfer \citep{Khan2006}. \clb{The strong magnetic field of HD\,125248 is expected to suppress all convective motions in the atmosphere, therefore we set the micro- and macroturbulent velocities to zero in all calculations.}

We adopted \cla{a radial} magnetic field with 7.2\,kG strength by phase-averaging the mean field modulus of HD\,125248, calculated from the dipole plus quadrupole models obtained by \citet{Bagnulo2002p1023}. The adopted estimate is close to the 6.6\,kG value predicted by the model of \citet{Glagolevskij2007p244}. The VALD database was used as a source of atomic data \citep{Kupka1999p119}. We used this  model atmosphere and estimated the abundances of a number of light elements as well as Fe-peak and rare earth elements. For this purpose we fitted synthetic profiles calculated with the {\sc SYNMAST} code \citep{Kochukhov10p5} to the phase-averaged intensity (Stokes~$I$) spectrum of HD\,125248. 

The abundances determined in this step were then used to compute a grid of model atmospheres and SEDs with $T_\mathrm{eff} =$~9000--10200\,K and $\log g =$~3.9--4.3 \clb{in 200\,K steps for the effective temperature and 0.05\,dex steps for the surface gravity}. The calculated flux distributions were then compared to spectrophotometric observations in the optical range \citep{Adelman1989p221}, IUE Newly Extracted Spectra (IUE NES) in the UV obtained with large aperture, and near-IR photometric observations (2MASS fluxes). \clb{We adopted zero interstellar reddening for HD\,125248, following \citet{Lipski2008p481} who assumed that stars closer than 100\,pc do not experience significant interstellar absorption. We note, however, that their value of $d=90$\,pc results from a parallax of $\pi=11.08\pm0.91$\,mas from an earlier version of the Hipparcos catalog \citep{Perryman1997p}. The latest estimate for the parallax of HD\,125248 is $\pi = 9.80 \pm 0.67$\,mas \citep{Leeuwen07p653}, suggesting a distance of 102\,pc, which is still close to 100\,pc.}

\begin{figure*}
  \centering
  \includegraphics[width=\textwidth,page=1]{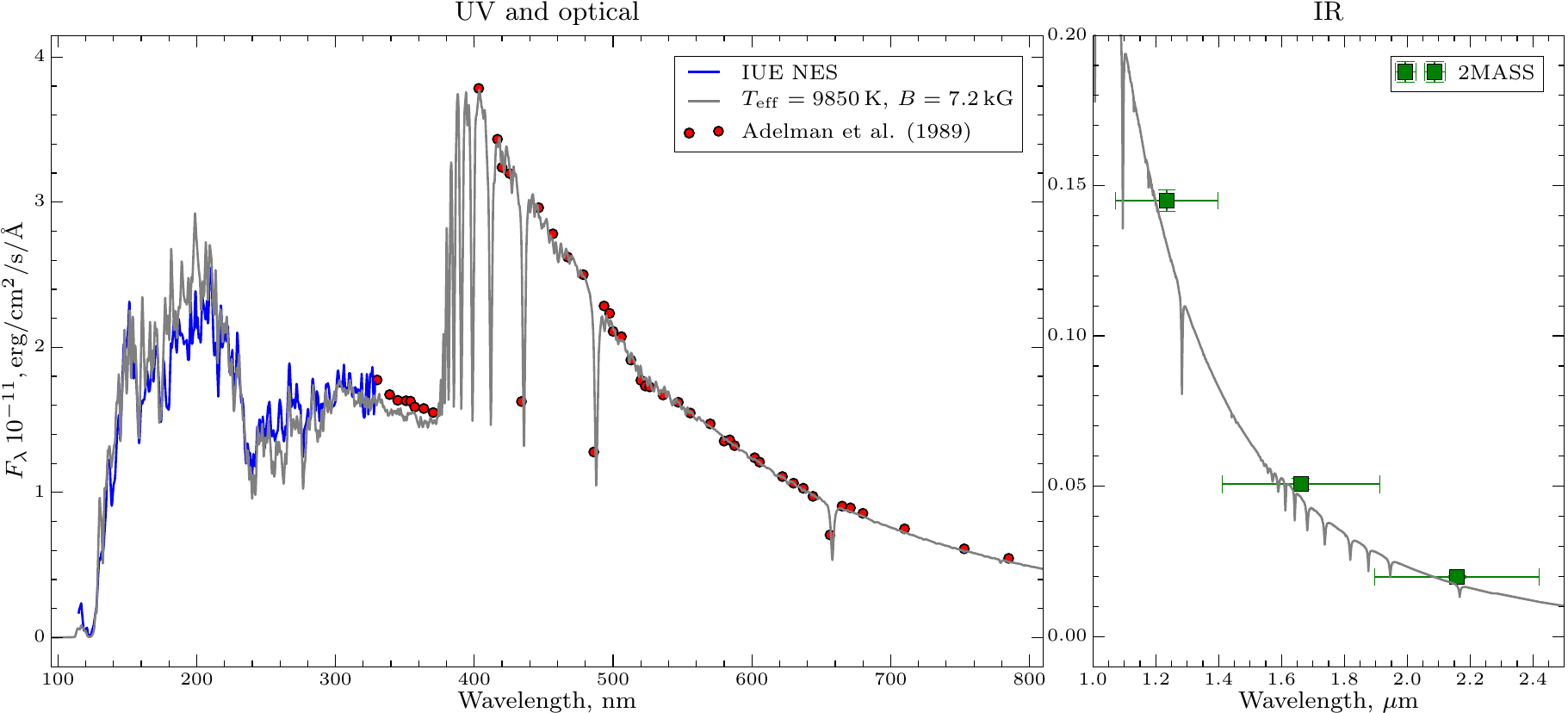}
  \caption{Comparison between the theoretical and observed spectrophotometry of HD\,125248. The light gray curve shows the spectral energy distribution computed using a model atmosphere with $T_\mathrm{eff}=9850$\,K, $\log g=4.05$, including the effects of individual non-solar abundances of HD\,125248 and \cla{a magnetic field} with strength $\langle B \rangle = 7200$\,G. The observations cover the UV (blue lines), optical (red circles) and near-IR (green squares) spectral regions.}\label{fig:sed}
\end{figure*}

The spectral energy distribution fitting procedure is illustrated in Figure~\ref{fig:sed}.
We found that an effective temperature of $T_\mathrm{eff} = 9850 \pm 250$\,K most accurately describes the slope of the Paschen continuum, the Balmer jump and the UV SED of HD\,125248. This value is fairly insensitive to the choice of surface gravity and agrees well with previous studies. Changing the effective temperature by 250\,K in each direction slightly improves the fit in the UV or near-IR region, therefore we consider this error bar to be a good estimate of the uncertainties of our fitting procedure.

\begin{figure}
  \centering
  \includegraphics[width=0.5\textwidth,page=2]{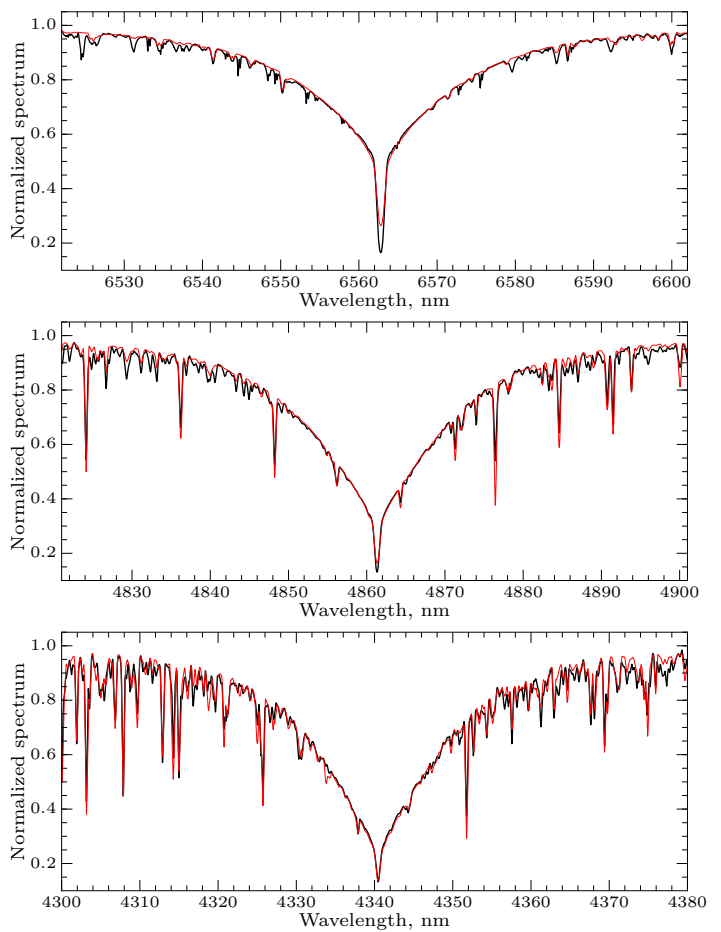}
  \caption{Comparison between the observed (thick black curve) and computed (thin red curve) profiles of the H$\alpha$, H$\beta$ and H$\gamma$ lines in the spectrum of HD\,125248.}
  \label{fig:hlines}
\end{figure}

Given the effective temperature determined in the previous step we constrained the surface gravity \cla{$\log g = 4.05\pm0.10$} by fitting synthetic profiles computed with {\sc SYNMAST} to the observed Balmer lines. The comparison between the observed and calculated profiles of H$\alpha$, H$\beta$ and H$\gamma$ is presented in Figure~\ref{fig:hlines}. The figure illustrates that the revised value of surface gravity describes the wings of the Balmer lines with good accuracy.

We combined the parallax of the star, $\pi = 9.80 \pm 0.67$\,mas, \citep{Leeuwen07p653} with its angular diameter, which was obtained in the SED fitting procedure, and derived the stellar radius $R=2.23\pm0.19\,R_\odot$. This value agrees well with the previous estimates by \citet[][$R=2.1\,R_\odot$]{Lipski2008p481} and \citet[][$R=1.97\,R_\odot$]{Bagnulo2002p1023}. \clb{The luminosity of HD\,125248 was estimated using the standard relation between the stellar radius and effective temperature, and was found to be $L=42.0\pm8.3\,L_\odot$, which is within one sigma in comparison to the value determined by \citet[][$L=31.6\pm5.8\,L_\odot$]{Kochukhov2006p763}.  The new values of the surface gravity and radius imply a mass $M=2.0\pm0.6\,M_\odot$, which is in good agreement with the stellar evolutionary model estimate by \citet[][$M=2.27\pm0.07\,M_\odot$]{Kochukhov2006p763}. Given the good agreement of the fundamental parameters between this study and the paper by \citet{Kochukhov2006p763} we deemed a redetermination of the age of HD\,125248 from evolutionary tracks to be unnecessary.}
 
Finally, the projected rotational velocity was estimated to be $v_e \sin i = 11.5\pm1.5$\,\cla{\kms{}} by fitting the Stokes~$I$ profiles of \ion{Fe}{i/ii} lines with low magnetic sensitivity. This value of $v_e \sin i$ agrees well with the determinations by \citet[][$v_e \sin i =10$\,\kms{}]{Abt1995p135} and \citet[][$v_e \sin i =9.7$\,\kms{}]{Mathys1995p746}.

The fundamental parameters of HD\,125248 are summarized in Table~\ref{tab:fundpars}.
\begin{table}
  \caption{Fundamental parameters of HD\,125248}
  \centering
  \begin{tabular}{lll}
  \hline
  \hline
  Parameter & Value & Reference\\
  \hline
$T_\mathrm{eff}$\,(K) & $9850 \pm 250$ & This study\\
$\log g$ & $4.05\pm0.10$ & This study\\
$L/L_\odot$ & $42.0\pm8.3$ & This study\\
$R/R_\odot$ & $2.23\pm0.19$ & This study\\
$M/M_\odot$ & $2.0\pm0.6$ & \clb{This study}\\
$\log(t)$\,(yr) & 8.37 (7.94--8.53) & \citet{Kochukhov2006p763}\\
  \hline
  \end{tabular}
  \label{tab:fundpars}
\end{table}

\subsection{Abundance analysis}\label{ssec:abund}
For the determination of accurate fundamental parameters of HD\,125248 it was necessary to determine the abundance of many chemical elements. For this purpose we estimated the abundance of most light elements, iron-peak and rare \cla{earth} elements as well as several s-process elements. The mean abundances are summarized in Table~\ref{tab:abund}. In short, the abundances of HD\,125248 appear to follow the general trend for other Ap stars --- several light elements are slightly deficient, the Fe-peak elements are overabundant by about \cla{one} dex with the exception of Cr, which is overabundant by about \cla{two} dex, while the rare earth elements are overabundant on average between two and three dex relative to their solar values. 

In order to be consistent with similar abundance and MDI studies of Ap stars we report abundances in $\log (N_X/N_\mathrm{tot})$ units. We note that in our analysis we did not account for hyperfine splitting or non-LTE effects, which can lead to systematic errors for certain elements. 
\begin{table}
\caption{Mean atmospheric abundances of the Ap star HD\,125248.}
\begin{tabular}{l|c|c|c|c}
\hline \hline
Ion & $\log(N_\mathrm{ion}/N_\mathrm{tot})$ & $n$ & $\log(N_\mathrm{el}/N_\mathrm{tot})$ & $\log(N/N_\mathrm{tot})_\odot$ \\
\hline
\ion{He}{i}& $ -1.80$ & 2 & $-$1.80 & $-$1.11 \\
\ion{O}{i}& $ -3.67 \pm 0.13$ & 2 & $-$3.67 & $-$3.35 \\
\ion{Mg}{ii} & $ -5.33$ & 1 & $-$5.33 & $-$4.44 \\
\ion{Si}{ii} & $ -4.22 \pm 0.13$ & 3 & $-$4.22 & $-$4.53 \\
\ion{Sc}{ii} & $ -8.32$ & 1 & $-$8.32 & $-$8.89 \\
\ion{Ti}{ii} & $ -6.54 \pm 0.18$ & 7 & $-$6.54 & $-$7.09 \\
\ion{V}{ii} & $ -7.57 \pm 0.30$ & 5 & $-$7.57 & $-$8.11 \\
\ion{Cr}{ii} & $ -4.58 \pm 0.45$ & 25 & $-$4.58 & $-$6.40 \\
\hline 
\ion{Mn}{i} & $ -5.86 \pm 0.11$ & 4 & \multirow{2}{*}{$-$5.75} & \multirow{2}{*}{$-$6.61} \\
\ion{Mn}{ii} & $ -5.42 \pm 0.19$ & 9 & & \\
\hline 
\ion{Fe}{i} & $ -3.42 \pm 0.17$ & 39 & \multirow{2}{*}{$-$3.39} & \multirow{2}{*}{$-$4.54} \\
\ion{Fe}{ii} & $ -3.30 \pm 0.33$ & 49 & & \\
\hline 
\ion{Co}{ii} & $ -5.57 \pm 0.22$ & 4 & $-$5.57 & $-$7.05 \\
\ion{Ni}{i} & $ -5.70$ & 1 & $-$5.70 & $-$5.82 \\
\ion{Sr}{ii} & $ -8.25 $ & 1 & $-$8.25 & $-$9.17 \\
\ion{Y}{ii} & $ -8.42 \pm 0.18$ & 2 & $-$8.42 & $-$9.83 \\
\ion{Zr}{ii} & $ -7.97 \pm 0.31$ & 6 & $-$7.97 & $-$9.46 \\
\ion{La}{ii} & $ -8.02 \pm 0.13$ & 4 & $-$8.02 & $-$10.94 \\
\ion{Ce}{ii} & $ -6.85 \pm 0.24$ & 28 & $-$6.85 & $-$10.46 \\
\hline 
\ion{Pr}{ii} & $ -7.47 \pm 0.44$ & 5 & \multirow{2}{*}{$-$7.59} & \multirow{2}{*}{$-$11.32} \\
\ion{Pr}{iii} & $ -7.64 \pm 0.27$ & 8 & & \\
\hline 
\ion{Nd}{ii} & $ -7.30 \pm 0.42$ & 2 & \multirow{2}{*}{$-$7.49} & \multirow{2}{*}{$-$10.62} \\
\ion{Nd}{iii} & $ -7.52 \pm 0.17$ & 13 &  &  \\
\hline 
\ion{Sm}{ii} & $ -7.65 \pm 0.13$ & 6 & $-$7.65 & $-$11.08 \\
\hline 
\ion{Eu}{ii} & $ -7.24 \pm 0.24$ & 7 & \multirow{2}{*}{$-$7.24} & \multirow{2}{*}{$-$11.52} \\
\ion{Eu}{iii} & $ -5.85 \pm 0.14$ & 2 &  & \\
\hline 
\ion{Gd}{ii} & $ -7.44 \pm 0.24$ & 16 & $-$7.44 & $-$10.97 \\
\ion{Tb}{iii} & $ -8.41 \pm 0.16$ & 3 & $-$8.41 & $-$11.77 \\
\ion{Dy}{ii} & $ -8.00 \pm 0.71$ & 2 & $-$8.00 & $-$10.94 \\
\ion{Ho}{iii} & $ -8.35 \pm 0.10$ & 4 & $-$8.35 & $-$11.56 \\
\hline 
\ion{Er}{ii} & $ -8.25 \pm 0.07$ & 2 & \multirow{2}{*}{$-$8.37} & \multirow{2}{*}{$-$11.12} \\
\ion{Er}{iii} & $ -8.48 $ & 1 &  & \\
\hline 
\ion{Lu}{ii} & $ -8.70 \pm 0.25$ & 2 & $-$8.70 & $-$11.94 \\
\hline
\end{tabular}
\label{tab:abund}
\tablefoot{The first \cla{column} identifies the ions for which we estimated the abundance given the second column. The number of lines analyzed for each species is given in the third column. The adopted abundances for the corresponding chemical elements are given in column four. The solar abundances reported by \citet{Asplund2009p481} are given in the last column for comparison.}
\end{table}

\subsubsection{Light elements: He to Ca}\label{sssec:abund:light}
The He abundance was obtained from two groups of \ion{He}{i} lines at $\lambda\,$4471\,\AA{} and  $\lambda$\,5875\,\AA{}. A calculation assuming solar helium abundance produces He lines that appear too deep contrary to what is observed in the mean spectrum of HD\,125248. Therefore, we adopted  abundance \abn{He}{-1.80}, which is smaller by around five times compared to the solar value. For the CNO-elements we could only measure  the abundance of O, which appears to be slightly deficient by 0.3\,dex. Abundances of other light elements are mostly solar, with the exception of Mg and Si. The former is strongly depleted by about 0.8\,dex, while the latter is slightly overabundant by 0.3\,dex.

\subsubsection{Iron peak elements: Sc to Ni}\label{sssec:abund:fepeak}
The abundance of Sc was inferred from the \ion{Sc}{i}\,4415\,\AA{} line, which requires \abn{Sc}{-8.32} to reproduce the line profile. The Ti abundance was straightforward to infer, thanks to the seven \ion{Ti}{ii} lines that we could successfully reproduce with \abn{Ti}{-6.54}. Vanadium, in contrast to Ti, had a few blended \ion{V}{ii} lines, which yielded an abundance higher by 0.5\,dex relative to the solar value. However, the large error makes this result somewhat inconclusive. We derived the Mn abundance from \ion{Mn}{i} and \ion{Mn}{ii} lines. Both estimates indicate that Mn is overabundant, with the abundance for \ion{Mn}{ii} being 0.44\,dex higher than the one for \ion{Mn}{i}. It is possible that this discrepancy is caused by hyperfine splitting. Cobalt has an overabundance of 1.55\,dex, which might be caused by lack of accurate oscillator strengths for \ion{Co}{ii} lines, or possibly, ignoring the hyperfine structure. Nickel, on the other hand, had only one \ion{Ni}{I}\,$\lambda$\,4980.166\,\AA{} line that wasn't distorted by blends. An attempt to measure the Ni abundance from this line gave an estimate that is only 0.1\,dex higher than the solar value.  The attempt to measure the Cr abundance resulted in \abn{Cr}{-4.58}, which is by $\sim$2\,dex higher than in the Sun.

\begin{figure}
  \centering
  \includegraphics{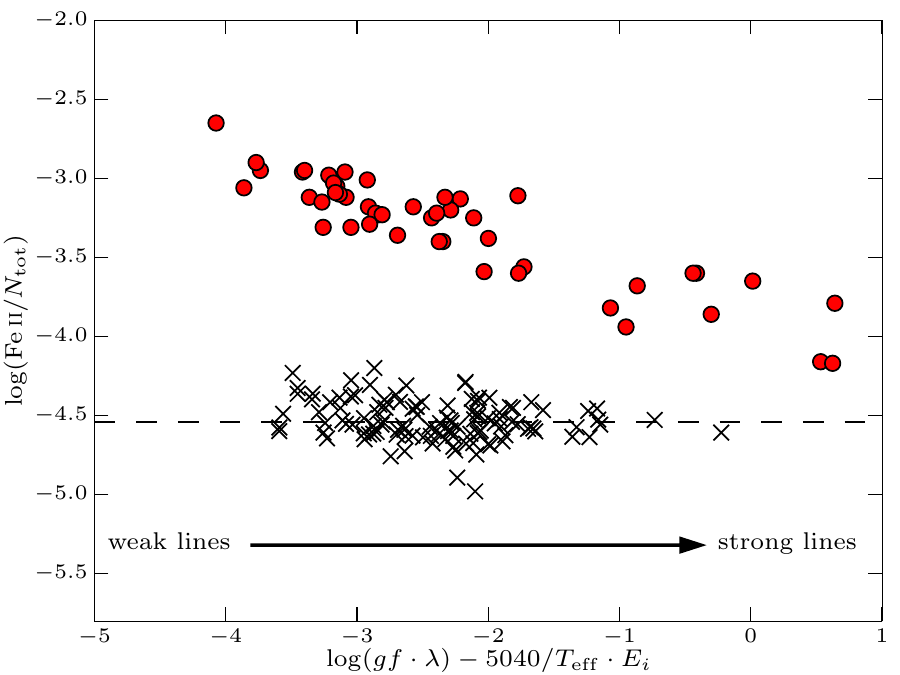}
  \caption{\cla{Individual} \ion{Fe}{ii} abundances as a function of line intensity in the atmospheres of HD\,125248 and 21~Peg. The filled circles represent  \ion{Fe}{ii} abundances for HD\,125248. The data for 21~Peg are plotted using ``x'' symbols. The solar Fe abundance is shown with a dashed line. }
  \label{fig:Fe-slope}
\end{figure}

Considering the importance of Fe in stellar atmospheres calculations, we took special care to measure its abundance from a large number of \ion{Fe}{i} and \ion{Fe}{ii} lines with excitation energy in the range $1<E_i<12$\,eV. The measurements from \ion{Fe}{i} and \ion{Fe}{ii} appear to agree well within the error bars, with only 0.14\,dex difference between them. The larger scatter of the \ion{Fe}{ii}-based abundance led us to explore further the relationship between the abundance measured from individual lines versus the line strength expressed as a function of the oscillator strength and excitation energy \citep[see, e.g.,][]{Ryabchikova2014p220}. We illustrate this in Fig.~\ref{fig:Fe-slope} for the \ion{Fe}{ii} lines. For comparison we added data for the normal A-type star 21~Peg 
\citep[Table 9 of][]{Fossati2009p945} that has an effective temperature similar to HD\,125248. As we can see from the figure, there is a visible trend of abundance with the line strength --- strong lines that are formed higher in the atmosphere give smaller individual abundances compared to the weak lines, which are formed near the photosphere. The presence of such a slope in abundance analysis of normal stars indicates that the effective temperature has been determined incorrectly. In the case of HD\,125248, it is impossible to remove this trend by correcting the effective temperature without contradicting the spectrophotometry of this star. The same technique was applied to the individual abundances of \ion{Cr}{ii} lines, which have a very significant 0.45\,dex scatter around the mean value, in spite of our best efforts to select mostly unblended Cr lines with various excitation energy and $\log gf$ factors. The analysis yielded similar results --- the abundances from individual \ion{Cr}{ii} lines show a strong dependence on the line strength. We consider these results to be a clear sign of vertical stratification of chemical elements. Detailed investigation of vertical abundance stratification is beyond the scope of our paper. 

\subsubsection{Neutron capture elements: Sr, Y, and Zr}\label{sssec:abund:neutron}
The Sr abundance was determined from the \ion{Sr}{ii} $\lambda$\,4215\,\AA{} line to be enhanced by 0.9\,dex compared to the solar value. We measured the Y abundance from two mostly unblended \ion{Y}{ii} lines, which require almost 1.5\,dex overabundance to properly reproduce the observed profiles in the mean spectrum. The zirconium abundance was reliably deduced from six \ion{Zr}{ii} lines to be also overabundant by 1.5\,dex. Our attempts at abundance analysis of other elements from the fifth group of the periodic table of elements were fruitless as we could not find spectral lines suitable for accurate abundance determination.

\subsubsection{Rare earth elements: La to Lu}\label{sssec:abund:ree}
The \cla{rare earth elements} appear to be significantly enhanced by $\sim$3\,dex and more compared to the solar values.

For some \cla{rare earth elements} it was possible to estimate their abundance using lines of the first and second ionization stages. These estimates demonstrated a very good agreement between abundances derived from the lines of different ions of the same element. The only exception to this rule is Eu, which shows significantly different values when determined from \ion{Eu}{ii} and \ion{Eu}{iii} lines. The Eu abundance derived from \ion{Eu}{ii} lines is similar to other \cla{rare earth elements}, while the \ion{Eu}{iii} abundance is higher by 1.4\,dex. The \ion{Eu}{iii} abundance, unfortunately, is less constrained because we could only determine it from the \ion{Eu}{iii} $\lambda$\,5376\,\AA{} and \ion{Eu}{iii} $\lambda$\,6666\,\AA{} lines. Note, that a similar pattern with the Eu-anomaly and the absence of other \cla{rare earth anomalies} has been observed in the atmosphere of another Ap star HD~144897 \citep{Ryabchikova2006p329} although more lines of \ion{Eu}{iii} were used in abundance analysis of this star. The lack of reliable atomic data (hfs constants, isotopic shifts, collision rates) for \ion{Eu}{iii} does not allow us to conclude if this discrepancy is caused by 
ignoring the hyperfine splitting and non-LTE effects \citep{Mashonkina2002p34} or by an actual physical mechanism that produces abundance differences between ionization states of this particular \cla{element}, \cla{atomic} diffusion, for example.
 
\section{Integral magnetic observables}\label{sec:mag}
The mean characteristics of a stellar magnetic field  such as the mean longitudinal field can be inferred directly from spectropolarimetric observations \citep[e.g.,][]{Landstreet2000p213,Bagnulo2002p1023}. Such measurements of the magnetic observables for HD\,125248 from polarization signals in individual lines are difficult because its spectrum is highly complex --- most spectral line profiles are severely blended by an amount that changes significantly with rotational phase. Therefore, we employed the least-squares deconvolution technique \citep[LSD,][]{Donati97p658} to the Stokes spectra of HD\,125248. This technique assumes that a stellar spectrum can be represented as a superposition of \clb{similar profiles} scaled by a factor that depends on the line strength, the magnetic sensitivity and the central wavelength. For this study, we used the multiprofile version of the LSD method introduced by \citet{Kochukhov10p5} that allows us to disentangle the mean profiles of a given set of elements while minimizing blending of the lines of other elements.

We used the VALD3 database  \citep{Ryabchikova2015p54005} as a source of atomic data necessary for construction of the LSD line mask, together with the model atmosphere and abundance table produced in Sect.~\ref{sec:fund}. After removing spectral lines affected by hydrogen and telluric lines, and lines with central depth less than 0.1 of the continuum intensity, we were left with a final line mask comprising 4177 lines\clb{, of them 1961 lines belong to Fe-peak elements, and 2122 lines belong to rare earth elements}. The final line mask is dominated by \ion{Fe}{ii} (861 lines) and \ion{Ce}{ii} (977 lines). 

\begin{figure}
  \centering
  \includegraphics[scale=0.98]{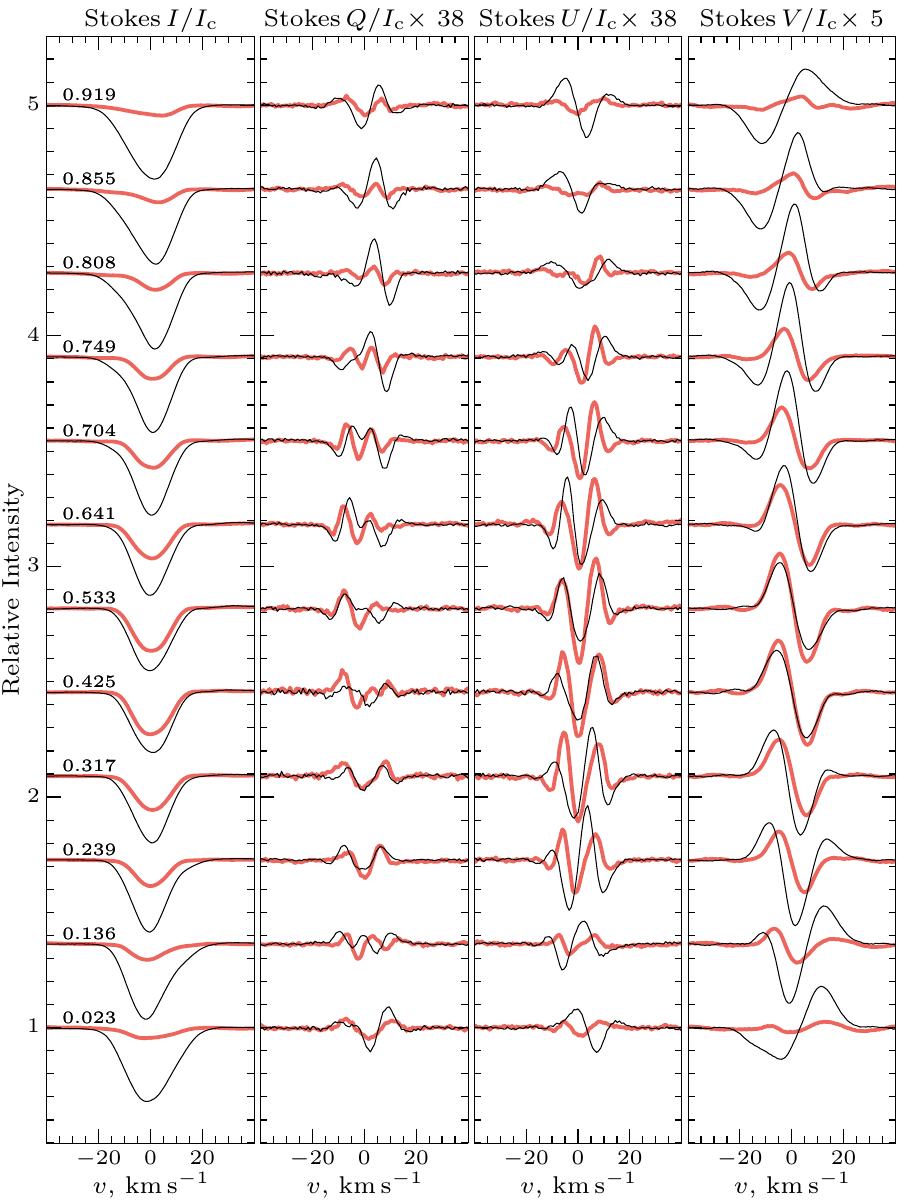}
  \caption{Stokes~$I$ (\textit{first panel}), $Q$ (\textit{second panel}), $U$ (\textit{third panel}), and $V$ (\textit{fourth panel}) LSD profiles of HD\,125248. The spectra are shifted vertically according to the rotational phases indicated in the first panel. The Stokes~$QU$ profiles have been rescaled by a factor of 38, and the Stokes~$V$ profiles by a factor of 5 to match the amplitude of the Stokes~$I$ profiles. The LSD profiles of Fe-peak elements are illustrated with thin lines (black) while the LSD profiles of the rare earth elements are shown with thick (red) lines.}
  \label{fig:lsdprofiles}
\end{figure}

The multiprofile LSD technique was applied to all observations of HD\,125248 to obtain LSD profiles of Fe-peak and rare earth elements (Fig.~\ref{fig:lsdprofiles}). In addition, we computed mean LSD profiles for the entire mask. All LSD profiles were calculated using the following normalization coefficients: $\lambda_0 = \SI{5000}{\angstrom}$, $d_0 = 1$, and $\bar{g}_0 = 1$.

The resulting LSD profiles for the Fe-peak and rare earth elements (Fig.~\ref{fig:lsdprofiles}) were compared to the profiles of several individual lines. This comparison showed that the resulting mean profiles properly describe the behavior with rotational phase of spectral lines of these species. \clb{The visual analysis of the LSD profiles shows what appears to be an almost total absence of rare earth element lines for phases around 0.0, when the lines of Fe-peak elements are strongest.} Interpretation of this unusual variation with rotational phase of individual lines of Fe-peak and rare earth elements is the goal of our magnetic Doppler imaging analysis, presented in Sect.~\ref{sec:mdi}.

We used the LSD profiles and calculated the longitudinal magnetic field, \bz{}, and net linear polarization $P_Q$ and $P_U$. The former is a useful measurement of the line-of-sight component of the magnetic field. It can be calculated from the first order moment of the Stokes~$V$ profiles \citep{Kochukhov10p5}. This is the most commonly used magnetic observable for early type stars. The net linear polarization is calculated from the equivalent width of the LSD Stokes~$Q$ and $U$ profiles and contains information about the transverse component of the magnetic field similar to what is diagnosed by the broadband linear polarization \citep{Leroy1995p79}. However, direct comparison between the two types of measurements is not possible unless they are scaled and shifted \citep[e.g.,][]{Wade00p823,Rusomarov2013p8}.

The longitudinal field and net linear polarization measurements are presented in Table~\ref{tab:int-meas}. The calculations were done for Fe-peak, rare earth elements and for the LSD profiles obtained with the full mask. The measurement uncertainties were obtained from the actual LSD profiles using the standard error propagation principle. In the same way we calculated the equivalent quantities from the null profiles, which were derived by applying the line mask and the LSD technique to the null Stokes spectra. These quantities are excellent indicators of spurious polarization signals. Our analysis of these quantities showed that any spurious contributions to the measurements presented in this section are well below the uncertainties due to the finite signal-to-noise ratio of the observations. 
\begin{table*}[!t]
  \tiny
  \caption{Integral magnetic field measurements obtained from the LSD Stokes profiles of HD\,125248.}
  \centering
  \resizebox{\textwidth}{!}{
  \begin{tabular}{ccrrrrrrrrrrrr}
  \hline
  \hline
    && \multicolumn{3}{c}{$\langle B_\mathrm{z} \rangle \pm \sigma_{\langle B_\mathrm{z} \rangle}\,(\mathrm{G})$} &\multicolumn{3}{c}{null $\langle B_\mathrm{z} \rangle \pm \sigma_{\langle B_\mathrm{z} \rangle}\,(\mathrm{G})$}&\multicolumn{3}{c}{$P_Q \pm \sigma_{P_Q}\,(\times10^{-4})$} & \multicolumn{3}{c}{$P_U \pm \sigma_{P_U}\,(\times10^{-4})$}\\
    HJD\,(2\,455\,000+) & Phase & \multicolumn{1}{c}{full} &  \multicolumn{1}{c}{Fe-peak} &  \multicolumn{1}{c}{REE} &  \multicolumn{1}{c}{full} &  \multicolumn{1}{c}{Fe-peak} &  \multicolumn{1}{c}{REE} &  \multicolumn{1}{c}{full} &  \multicolumn{1}{c}{Fe-peak} &  \multicolumn{1}{c}{REE} \\
\hline
1017.77255 & 0.023 & $-1854\pm25$ &$-1811\pm20$ &$-553\pm164$ & $ -5\pm4$ & $ -6\pm4$ & $ -4\pm30$ &$ -1.5\pm1.1$ &$  0.3\pm0.9$ &$-15.3\pm8.4$ &$  8.8\pm1.1$ &$  5.0\pm0.8$ &$ 24.0\pm7.6$\\
1018.81486 & 0.136 & $-1052\pm24$ &$-1180\pm19$ &$ 893\pm116$ & $ -6\pm5$ & $ -8\pm4$ & $ -2\pm27$ &$ 19.7\pm1.0$ &$ 21.4\pm0.8$ &$ -8.7\pm5.7$ &$-12.7\pm1.0$ &$ -8.1\pm0.8$ &$-33.1\pm5.3$\\
1019.77340 & 0.239 & $  571\pm20$ &$  260\pm17$ &$1946\pm 64$ & $-11\pm5$ & $-12\pm4$ & $ -7\pm16$ &$  2.8\pm1.1$ &$ 13.2\pm0.9$ &$-36.1\pm3.9$ &$-31.2\pm1.4$ &$-38.0\pm1.2$ &$  7.0\pm4.8$\\
1345.84686 & 0.317 & $ 1612\pm23$ &$ 1307\pm23$ &$2249\pm 56$ & $ -2\pm6$ & $ 11\pm6$ & $-28\pm15$ &$  2.3\pm1.6$ &$ -8.2\pm1.5$ &$ 35.3\pm4.1$ &$-22.1\pm1.6$ &$-24.9\pm1.5$ &$-10.3\pm4.0$\\
1346.85386 & 0.425 & $ 1970\pm30$ &$ 1584\pm33$ &$2579\pm 61$ & $ -8\pm6$ & $ -5\pm6$ & $-22\pm11$ &$  9.8\pm2.1$ &$ -8.4\pm2.2$ &$ 45.0\pm4.4$ &$ 17.7\pm1.6$ &$ 12.9\pm1.7$ &$ 26.2\pm3.4$\\
1347.85514 & 0.533 & $ 1917\pm28$ &$ 1560\pm31$ &$2670\pm 58$ & $  6\pm5$ & $  0\pm5$ & $ 13\pm10$ &$-17.1\pm1.3$ &$-13.4\pm1.3$ &$-24.2\pm2.7$ &$  9.9\pm1.4$ &$  6.0\pm1.4$ &$ 15.5\pm2.9$\\
1348.86017 & 0.641 & $ 1497\pm21$ &$ 1223\pm21$ &$2256\pm 55$ & $ 11\pm4$ & $ 13\pm4$ & $ 18\pm11$ &$-14.9\pm1.2$ &$-16.0\pm1.1$ &$-13.2\pm3.2$ &$ 11.7\pm1.2$ &$  0.8\pm1.2$ &$ 42.5\pm3.3$\\
1014.79860 & 0.704 & $  807\pm19$ &$  529\pm17$ &$1815\pm 59$ & $ 14\pm4$ & $ 15\pm4$ & $ 11\pm12$ &$-20.2\pm1.1$ &$-15.6\pm1.0$ &$-36.9\pm3.9$ &$  3.1\pm1.1$ &$ 15.7\pm1.0$ &$-43.0\pm4.0$\\
1349.86061 & 0.749 & $  135\pm20$ &$ -111\pm16$ &$1371\pm 76$ & $ 12\pm4$ & $ 17\pm4$ & $-30\pm18$ &$-23.4\pm1.1$ &$-20.2\pm0.9$ &$-35.6\pm4.6$ &$  6.7\pm1.0$ &$ 12.3\pm0.8$ &$-22.7\pm4.2$\\
1015.77181 & 0.808 & $ -780\pm23$ &$ -958\pm18$ &$ 601\pm106$ & $ 18\pm5$ & $ 17\pm4$ & $ 15\pm24$ &$-26.7\pm1.2$ &$-21.3\pm1.0$ &$-52.6\pm6.3$ &$ 14.3\pm1.0$ &$  9.8\pm0.8$ &$ 39.5\pm5.4$\\
1350.85141 & 0.855 & $-1339\pm24$ &$-1428\pm19$ &$ -32\pm130$ & $ 15\pm4$ & $  9\pm4$ & $ 58\pm24$ &$-16.7\pm1.1$ &$-12.1\pm0.8$ &$-39.2\pm6.5$ &$ -0.1\pm1.0$ &$ 10.3\pm0.8$ &$-74.7\pm6.1$\\
1016.80012 & 0.919 & $-1794\pm25$ &$-1794\pm19$ &$-523\pm155$ & $ -4\pm4$ & $ -2\pm3$ & $-27\pm25$ &$-12.4\pm1.0$ &$-12.0\pm0.8$ &$ -5.5\pm7.2$ &$ 15.2\pm1.1$ &$ 10.3\pm0.8$ &$ 49.4\pm7.5$\\
  \hline
  \end{tabular}}
  \label{tab:int-meas}
  \tablefoot{First and second columns list heliocentric JD and rotational phase, calculated according to the improved period (Sect.~\ref{ssec:rot}). Columns 3--5 provide \bz{} measurements for the Stokes $V$ LSD profiles obtained with the full mask, for the lines of Fe-peak elements, and for the REE lines. \clb{Columns 6--8 provide the \bz{} measurements from the null Stokes~$V$ profiles. Columns 9--11 report the Stokes~$Q$ net linear polarization measurements for the same three sets of LSD profiles, and columns 12--14 provide the net linear polarization measurements for the LSD Stokes~$U$ profiles.}}
\end{table*}

The full LSD line mask was also applied to the circular polarization measurements obtained at the NOT. We use the resulting mean longitudinal field measurements only to extend the baseline for the rotational period determination in Sect.~\ref{ssec:rot}. The \bz{} measurements from SOFIN spectra are presented in Table~\ref{tab:bz-NOT}.
\begin{table}
\caption{Mean longitudinal magnetic field of HD\,125248 obtained from NOT observations.}
\begin{tabular}{crr}
\hline
\hline
HJD & Phase & $\langle B_\mathrm{z} \rangle$\,(G)\\
\hline
2452064.44814 & 0.733 & $  274 \pm 108$\\
2452065.46507 & 0.842 & $-1316 \pm 122$\\
2452066.46713 & 0.950 & $-2125 \pm 125$\\
2452067.46349 & 0.057 & $-1900 \pm 121$\\
2452068.47056 & 0.165 & $ -698 \pm 107$\\
2452069.46535 & 0.272 & $ 1144 \pm 108$\\
2452070.46415 & 0.380 & $ 1940 \pm 143$\\
2452071.46090 & 0.487 & $ 2110 \pm 157$\\
2452072.44919 & 0.593 & $ 1815 \pm 132$\\
\hline
\end{tabular}\label{tab:bz-NOT}
\tablefoot{These measurements were derived from the LSD profiles calculated using the full mask.}
\end{table}

\subsection{Rotational period}\label{ssec:rot}
Given a large number of magnetic field studies of HD\,125248 it is possible to find \cla{a} highly precise rotational period. The analyzed list of longitudinal magnetic field measurements contains all published data with the exception of measurements produced only from \cla{rare earth element lines}. We also did not include the early measurements by \citet{Babcock1951p1} as they appeared to be too inaccurate for our study. The data by \citet{Babcock1958p141} were also discarded as this author did not provide accurate enough observation dates. Our list of \bz{} measurements includes the \ion{Fe}{i}, \ion{Fe}{ii}, and \ion{Ti}{ii} estimates by \citet{Hockey1969p543}, measurements of \ion{Fe}{ii} lines by \citet{Mathys1994p547} and \citet{Mathys97p475}, as well as data by \citet{Leone2001p118} and \citet{Shorlin2002p637}. \cla{To this list} we also added the photopolarimetric longitudinal field measurements from the wings of H$\beta$ lines by \citet{Landstreet1975p624} and \citet{Borra1980p421}. However, we rescaled these measurements by a factor of 0.85 to better match the observed \bz{} amplitude from the Fe-peak elements. These data were finally complemented by our own longitudinal field measurements from the recent HARPSpol observations and older NOT observations. 

We performed period search on the entire data set, comprised of 78 measurements in total, spanning over a time period of more than \cla{40 years}. As a starting guess for the fitting procedure we used the last reported period, $P_\mathrm{rot} = 9.29545$\,d, derived by \citet{Leone2001p118}, and adopted the zero point from their paper. The best fit curve was derived using four frequency components, which is necessary to reproduce the observed variations of the longitudinal field measurements. With the zero point from \citet{Leone2001p118} and our improved stellar rotational period the final ephemeris is given by
$$
\mathrm{HJD}({\langle B_\mathrm{z} \rangle}_\mathrm{min}) = 2\,433\,103.95 + 9.29558(6)\cdot E.
$$

\begin{figure}
\centering
\includegraphics[width=0.49\textwidth]{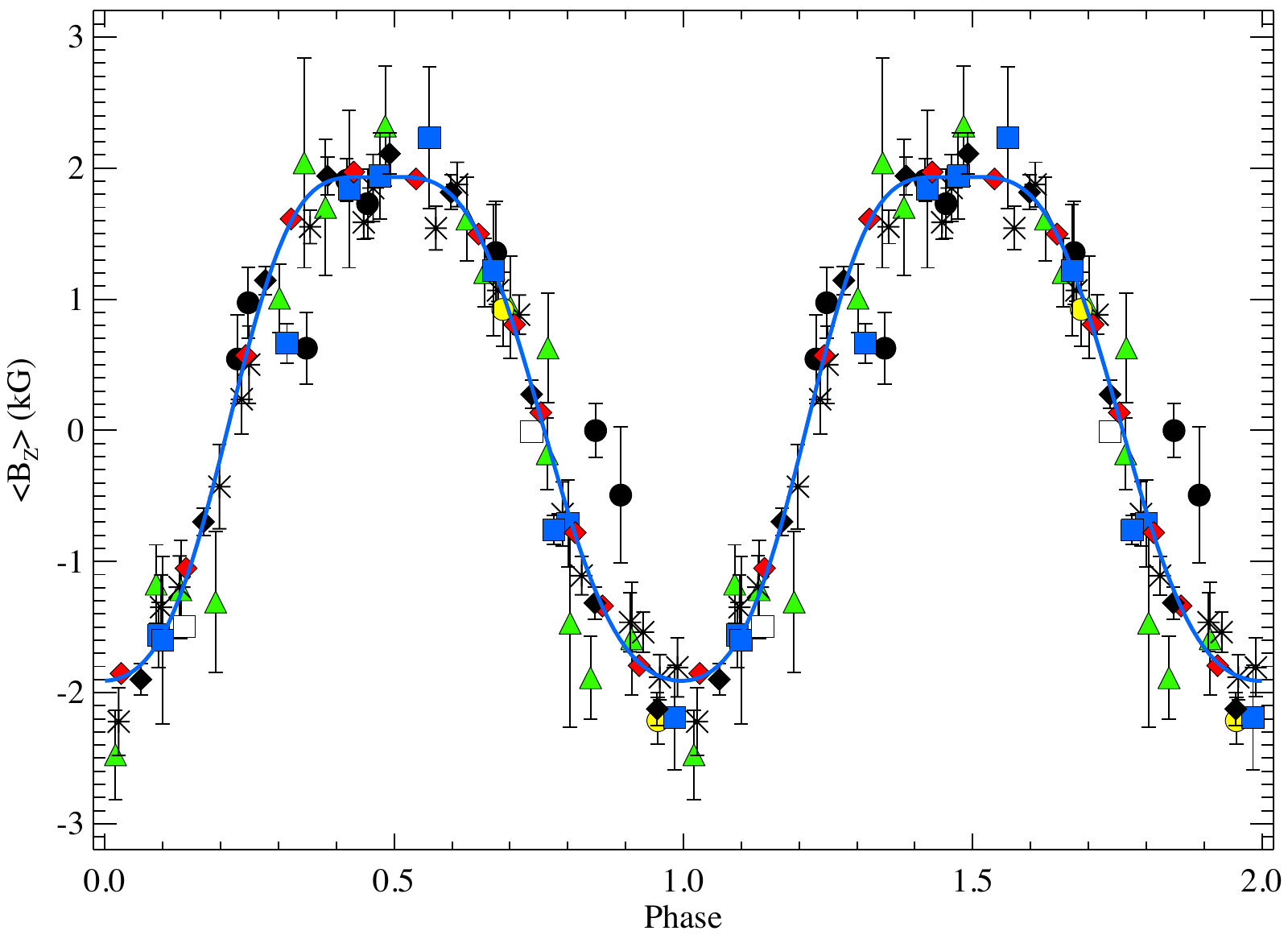}
\caption{\label{fig:period}Variability of the longitudinal magnetic field of HD\,125248 with rotational phase. The symbols correspond to the following data sets: HARPSpol (red diamonds), NOT (black diamonds), \citet{Hockey1969p543} (black bullets), \citet{Landstreet1975p624} and \citet{Borra1980p421} (green triangles), \citet{Mathys1994p547} (asteriks), \citet{Mathys97p475} (yellow bullets), \citet{Leone2001p118} (blue squares), \citet{Shorlin2002p637} (white squares). The solid thick curve is a fourth-order Fourier fit to the longitudinal field measurements.}
\end{figure}

Figure~\ref{fig:period} shows the variations of the mean longitudinal field measurements of HD\,125248 phased according to the revised rotational period. All phases given in our paper were computed according to the ephemeris given above. 

\section{Magnetic Doppler \cla{imaging}}\label{sec:mdi}
\subsection{Methodology}\label{ssec:mdi:methods}
The availability of high-quality spectropolarimetric observations in all four Stokes parameters for  \cla{HD\,125248 gives us the opportunity} to carry out a detailed tomographic reconstruction of its magnetic field and chemical abundance distributions. Magnetic Doppler imaging (MDI) is a computational technique that attempts to fit a set of observed Stokes parameters with synthetic spectra by adjusting the surface distribution of the magnetic field and the abundance of one or more chemical elements. The surface distribution maps are adjusted iteratively until the synthetic spectra properly describe the behavior of the observed line profiles with rotational phase in all Stokes parameters. We performed this task with the MDI code \invers{10} introduced by \citet{Piskunov2002p736} and \citet{Kochukhov2002p868}. In this study we closely follow already established methodological practices that have been used in recent MDI studies \citep[e.g.,][]{Rusomarov2015p123,Kochukhov2015p79}.

In the case when we have a full Stokes~$IQUV$ vector of spectropolarimetric observations, we can find the surface abundance $\varepsilon$ and magnetic field distribution $\vec{B}$ maps by solving the following non-linear least squares minimization problem:
\begin{equation}\label{eq:mini}
\begin{aligned}
\Psi(\varepsilon, \vec{B}) =& \sum_{\varphi \lambda} w_I (I_{\varphi \lambda}^\mathrm{obs} - I_{\varphi \lambda}^\mathrm{calc}(\varepsilon, \vec{B}))^2 / \sigma_{I \varphi \lambda}^2 \\
+& \sum_{\varphi \lambda} w_Q  (Q_{\varphi \lambda}^\mathrm{obs} - Q_{\varphi \lambda}^\mathrm{calc}(\varepsilon, \vec{B}))^2 / \sigma_{Q \varphi \lambda}^2 \\
+& \sum_{\varphi \lambda} w_U (U_{\varphi \lambda}^\mathrm{obs} - U_{\varphi \lambda}^\mathrm{calc}(\varepsilon, \vec{B}))^2 / \sigma_{U \varphi \lambda}^2 \\
+& \sum_{\varphi \lambda} w_V  (V_{\varphi \lambda}^\mathrm{obs} - V_{\varphi \lambda}^\mathrm{calc}(\varepsilon, \vec{B}))^2 / \sigma_{V \varphi \lambda}^2 \\
+& \Lambda_a \cdot \mathcal{R}_a(\varepsilon) + \Lambda_f \cdot \mathcal{R}_f(\vec{B}) \rightarrow \min, \\
\end{aligned}
\end{equation}
where $\lambda$ and $\varphi$ are the wavelength and rotational phase of each spectral pixel, and $\sigma_{\lambda\varphi}$ \cla{denotes its relative error}. The functions $\mathcal{R}_a$ and $\mathcal{R}_f$ are the regularization functionals of the abundance and magnetic field distribution maps, and $\Lambda_a$ and $\Lambda_f$ are the respective regularization parameters. To ensure that the relative contribution to the total discrepancy function $\Psi$ of different Stokes parameters is comparable we introduce the weights $w$. The parameters $\Lambda_a$ and $\Lambda_f$ control the contribution of the regularization that is included into the total discrepancy function $\Psi$. In practice we do not allow the contribution of the regularization to become less than 10 percent of the total discrepancy function. This ensures that our maps reproduce the observations reasonably well without fitting them down to the noise level. 

In this study we use first order Tikhonov regularization \citep{Tikhonov1977p} for the abundance maps, and a penalty function of the form $\sum_{\ell,m} \ell^2(\alpha^2_{lm} + \beta^2_{lm} + \gamma^2_{lm})$ for the magnetic field \citep{Donati06p629,Kochukhov2014p83}. The Tikhonov regularization leads to the smoothest possible maps for the abundance distributions as permitted by the observations, the penalty function constrains the inversion so that the magnetic field map contains the least possible contribution from high-order harmonic modes. 

Special care needs to be taken when computing the synthetic spectra in Eq.~(\ref{eq:mini}). In order to model the polarization signatures in a realistic way we calculate the local Stokes profiles for a large number of points on the stellar surface, we then   convolve these profiles with a Gaussian to simulate the spectral resolution of the instrument, and Doppler shift the resulting profiles for each rotational phase $\varphi$. The local profiles are then integrated over the visible stellar disk for each phase $\varphi$ and wavelength point $\lambda$. Finally they are normalized by a phase-independent continuum intensity. The integration procedure in \invers{10} also accounts for the projected area of each surface element for each phase $\varphi$. 

An important characteristics of the \invers{10} code is the way we compute the local Stokes profiles. Instead of using simplifying approximations in the form of fixed local Gaussian profiles or Milne-Eddington atmospheres \citep[e.g.,][]{Donati1997p1135}, \invers{10} solves the equations of polarized radiative transfer numerically in a realistic model atmosphere for each point on the stellar surface. 

The version of the \invers{10} code used here represents the magnetic field as a superposition of poloidal and toroidal components, each expressed as a spherical harmonics series \citep{Donati06p629,Kochukhov2014p83}.  In this formalism, the field is represented by a set of harmonic coefficients $\alpha_{\ell m}$, $\beta_{\ell m}$, and $\gamma_{\ell m}$, representing the poloidal radial, the poloidal horizontal, and the toroidal horizontal components, respectively. The expansion of the magnetic field components is truncated at $\ell_\mathrm{max}$. For HD\,125248, given the relatively low projected rotational velocity, we found that an expansion up to $\ell_\mathrm{max} = 10$ is sufficient to describe the magnetic field. 

\subsection{Spectral line selection}\label{ssec:mdi:lines}
Simultaneous reconstruction of the surface magnetic field and abundance distributions of as many chemical elements as possible allows to achieve a robust MDI solution \citep{Kochukhov2002p868,Silvester2014p182}. However, these two goals have slightly different requirements when it comes to line selection. The strong magnetically sensitive lines are excellent for magnetic field mapping, but they are not optimal for abundance reconstruction because they may be saturated over a large area of the stellar surface. In contrast, the weaker lines are better suited for abundance mapping and for deriving the projected rotational velocity $v_e \sin i$, but due to low amplitude of polarization signals, are less suitable for magnetic field mapping if used by themselves. Therefore, a mix of lines with different magnetic sensitivity and line strength is necessary to achieve a robust reconstruction of the surface magnetic field and abundance distributions. An additional line selection criterion is also the absence of significant blending.

On the basis of these criteria we selected 12~\ion{Fe}{i/ii}, 7~\ion{Cr}{ii}, and three lines each for \ion{Ce}{ii}, \ion{Gd}{ii}, and \ion{Nd}{iii} ions. Some of the Fe and Cr lines, e.g., \ion{Fe}{ii} $\lambda$\,5018\,\AA{}, \ion{Cr}{ii} $\lambda$\,4824\,\AA{}, as well as \ion{Nd}{iii} $\lambda$\,5851\,\AA{} have already been used in previous MDI studies. Unfortunately, other well known \ion{Fe}{ii} lines with strong polarization signals are severely blended by other elements in the spectrum of HD\,125248 and could not be used in our MDI study. The line selection contains lines with excitation energy spanning a range of close to 10\,eV, and $\log gf$ factors that span 3.4\,dex. We used lines that have low magnetic sensitivity ($\bar{g} \le 1.0$) and lines that are strongly sensitive with large Land\'e factors ($\bar{g} \ge 1.5$).

The line list adopted for the MDI of HD\,125248 is presented in Table~\ref{tab:linelist}. The atomic data for these lines were extracted from the latest version of the VALD database \citep{Ryabchikova2015p54005}. As reported in Sect.~\ref{ssec:abund}, we found a definite trend \clb{between the relative strength and abundance} for spectral lines of the \ion{Fe}{ii} and \ion{Cr}{ii} ions. Such discrepancies have also been reported for other magnetic Ap stars and are linked to vertical stratification \citep{Ryabchikova2005p973,Ryabchikova2014p220}. Part of them may be caused by non-LTE effects.
Since neither is included in our MDI modeling we introduced additional corrections to the oscillator strengths of certain lines. These corrections were automatically calculated by the \invers{10} code.
\begin{table}
\caption{Spectral lines used in magnetic Doppler imaging of HD\,125248.}\label{tab:linelist}
\centering
\begin{tabular}{lcccc}
\hline
\hline
Ion & $\lambda$\,(\AA{})& $E_\mathrm{i}$\,(eV)& $\log gf$& $\bar{g}$\\
\hline
\ion{Ce}{ii}& 4460.207&  0.478&  0.280 & 0.690\\ %
\ion{Ce}{ii}& 4486.909&  0.295& $-$0.180 & 0.985\\ %
\ion{Ce}{ii}& 4562.359&  0.478&  0.210 & 1.028\\ %
\hline
\ion{Gd}{ii}& 4597.910&  0.602& $-$0.830& 1.585\\ %
\ion{Gd}{ii}& 4732.609&  1.102& $-$0.540& 1.555\\ %
\ion{Gd}{ii}& 5092.249&  1.727& $-$0.230& 1.535\\ %
\hline
\ion{Nd}{iii}& 5677.179&  0.631& $-$1.450& 1.600\\ %
\ion{Nd}{iii}& 5851.542&  0.461& $-$1.550& 1.660\\ %
\ion{Nd}{iii}& 6145.068&  0.296& $-$1.330& 0.690\\ %
\hline
\ion{Cr}{ii}& 4554.988&  4.071& $-$1.282& 1.330\\ %
\ion{Cr}{ii}& 4618.803&  4.074& $-$0.840& 0.918\\ %
\ion{Cr}{ii}& 4824.127&  3.871& $-$0.970& 1.340\\ %
\ion{Cr}{ii}& 5420.922&  3.758& $-$2.458& 1.490\\ %
\ion{Cr}{ii}& 6134.465&  6.578& $-2.285^*$ & 0.595\\ 
\ion{Cr}{ii}& 6138.721&  6.484& $-1.793^*$ & 0.915\\ 
\ion{Cr}{ii}& 6336.263&  4.073& $-$3.759& 0.665\\ %
\hline
\ion{Fe}{ii}& 4508.280&  2.856& $-$2.250& 0.500\\ %
\ion{Fe}{ii}& 4576.333&  2.844& $-$2.920& 1.185\\ %
\ion{Fe}{ii}& 5018.436&  2.891& $-$1.220& 1.935\\ %
\ion{Fe}{ii}& 5061.703& 10.308&  $0.606^*$& 1.360\\ 
\ion{Fe}{ii}& 5169.028&  2.891& $-$1.250& 1.325\\ %
\ion{Fe}{ii}& 5197.568&  3.230& $-$2.100& 0.660\\ %
\ion{Fe}{i}& 5615.644&  3.332&  $-0.050^*$ & 1.200\\ 
\ion{Fe}{ii}& 5961.706& 10.678&  $0.995^*$ & 1.192\\ 
\ion{Fe}{ii}& 5991.371&  3.153& $-$3.540 & 0.813\\ %
\ion{Fe}{ii}& 6084.102&  3.199& $-3.930^*$ & 0.712\\ 
\ion{Fe}{ii}& 6446.407&  6.223& $-$1.960& 1.240\\ %
\ion{Fe}{ii}& 6482.204&  6.219& $-1.803^*$ & 0.940\\ 
\hline
\end{tabular}
\tablefoot{The columns give the ion, central wavelength $\lambda$, excitation potential of the lower atomic level $E_i$, the oscillator strength $\log gf$, and the effective Land\'e factor $\bar{g}$ of the spectral lines used for the inversions. \\
$({}^*)$ $\log gf$ value automatically adjusted in the MDI inversion.}
\end{table}

\subsection{Optimization of $v_e \sin i$, and orientation of the rotation axis}\label{ssec:mdi:opt-params}
\cla{In addition to an accurate stellar atmosphere model and a complete line list, MDI also requires knowledge of the orientation of the stellar rotation axis and the projected rotational velocity $v_e \sin i$.} The orientation of the rotation axis is specified by the two angles $i$ and $\Theta$. The inclination $i$ is the tilt of the rotation axis relative to the observer's line of sight; it has values in the range $[0^\circ, 180^\circ]$. The values of $i$ between $0^\circ$ and $90^\circ$ correspond to the situation when the star rotates counterclockwise as seen by the observer. The azimuth angle $\Theta$ determines the orientation of the rotation axis in the plane on the sky; it has values in the range $[0^\circ, 360^\circ]$. However, because the Stokes~$QU$ parameters depend on the trigonometric functions of $2\Theta$ there is an inherent ambiguity between values of $\Theta$ and $180^\circ + \Theta$. 

\citet{Kochukhov2002p868} showed that incorrect values of these parameters lead to increase of the $\chi^2$ value for the fit to the observed Stokes profiles. An incorrect projected rotational velocity $v_e \sin i$ leads to worse fit for the Stokes~$I$ parameters; incorrect values for the $i$ and $\Theta$ parameters results in similarly worse fit for the Stokes~$Q$ and $U$ parameters. 

The initial value for the inclination $i=75^\circ \pm 25^\circ$ (or $i=105^\circ \pm 25^\circ$) was calculated from the oblique rotator relation taking into account the stellar radius $R=2.23\,R_\odot$ and the projected rotational velocity $v_e \sin i = 11.5 \pm 1.5$\,\kms{} estimated in Sect.~\ref{sec:fund}, and the rotational period $P_\mathrm{rot} = 9.29558$\,d found in Sect.~\ref{ssec:rot}.

The initial guess for the azimuthal angle $\Theta$ of the rotational axis was estimated in the following way. We fitted our longitudinal magnetic field and net linear polarization measurements calculated for the full mask (Sect.~\ref{sec:mag}, Table~\ref{tab:int-meas}) 
with predictions of the ``canonical'' model for a dipolar field geometry \citep{Landolfi1993p285}. We did not rescale the net linear polarization measurements to match the observed amplitude of the broadband linear polarization measurements \citep{Leroy1995p79} because there are only five such measurements for HD\,125248 with error bars comparable to the amplitude of the observations themselves. Besides, we only aimed to produce a rough estimate of the azimuthal angle $\Theta$ and investigate how it depends on $i$, keeping in mind that a dipole model is known to produce worse results than a more sophisticated combination of a dipole and quadrupole models \citep{Bagnulo1999p865}. With these considerations we found that for different values of $i = 75^\circ \pm 25^\circ$ the azimuthal angle was always close to $30^\circ$. This is the value that we adopted as an initial guess.

\begin{figure}
  \centering
  \includegraphics[width=0.4975\textwidth]{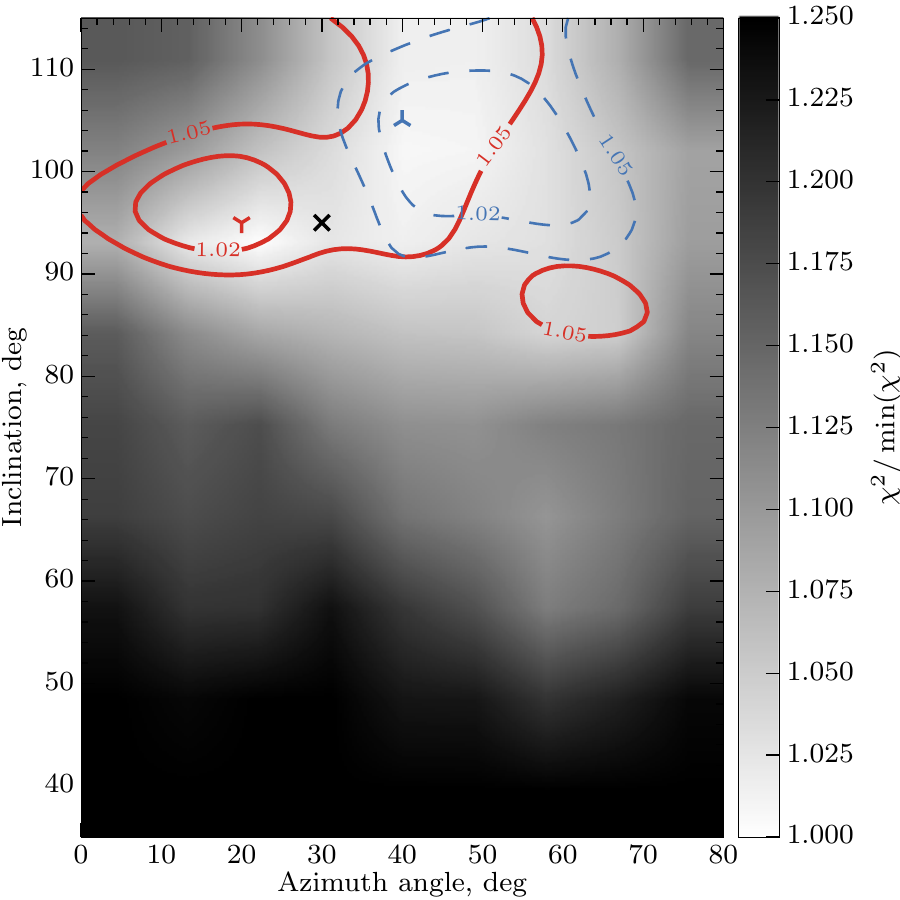}
  \caption{\clb{Variation of the $\chi^2$ of the fit to the Stokes~$QU$ profiles with inclination $i$ and azimuth angle $\Theta$.} The contours plotted with thick (red) lines show the 2\%, and 5\% increase from the minimum marked with a $\Ydown$ symbol for the Fe and Cr lines. The contours plotted with dashed (blue) lines show the same for the rare earth lines, the minimum in this case is marked with a $\Yup$ symbol. The $\times$ symbol indicates the adopted values of $i$ and $\Theta$.}\label{fig:mdi:angles}
\end{figure}

We then optimized the values of $i$ and $\Theta$ by calculating 81 MDI inversions on a grid $i \in [35^\circ, 115^\circ]$ and $\Theta \in [0^\circ, 80^\circ]$ with $10^\circ$ steps for both angles. The minimum value of the $\chi^2$ for the fit to the Stokes~$Q$ and $U$ profiles when considering all spectral lines was found at $i=95^\circ$ and $\Theta=20^\circ$. Furthermore, we inspected the position of the minima separately for Fe and Cr, and the rare earth elements. It was found out that the minimum for the combined set of Fe and Cr lines is close to the one derived for all lines. The rare earth element lines have minimum at $i=105^\circ$ and $\Theta=40^\circ$. Because we are simultaneously modeling both groups of lines we adopted $i=95^\circ$ and $\Theta=30^\circ$ and $10^\circ$ errors for both quantities. The goodness of fit for the Stokes~$QU$ parameters is shown in Fig.~\ref{fig:mdi:angles}.

\begin{figure}
  \centering
  \includegraphics[width=0.4975\textwidth]{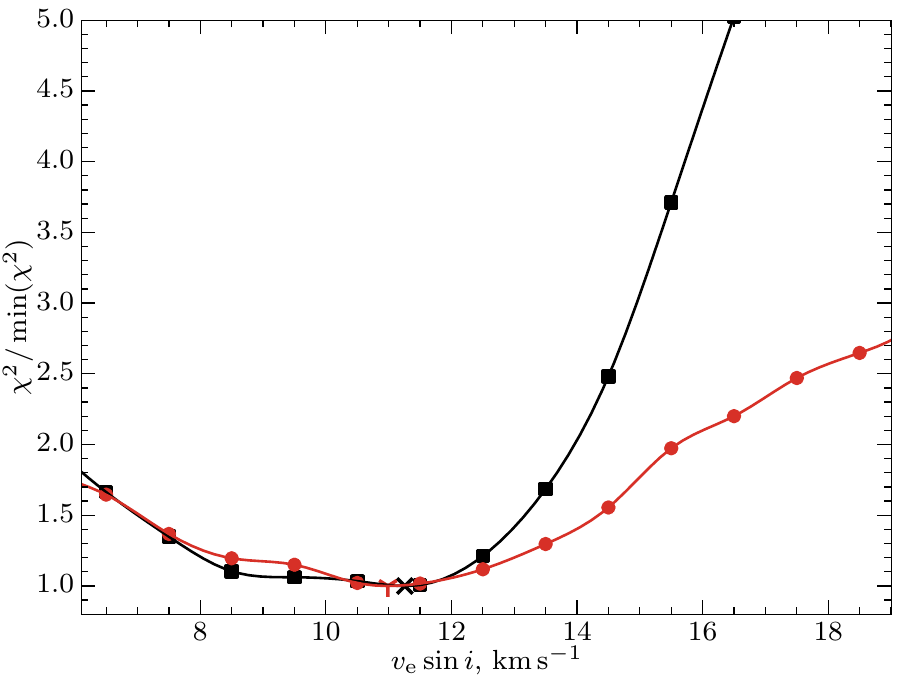}
  \caption{\clb{Variation of the $\chi^2$ of the fit to the Stokes~$I$ profiles as a function of the projected rotational velocity.} The curve marked with squares (black) shows the normalized deviation for the Fe and Cr lines; their minimum is marked with the $\times$ symbol. The curve marked with bullets (red) shows the same for the rare earth element lines; the minimum is marked with the $\Ydown$ symbol. The solid curves are interpolating cubic splines used to find the optimal projected rotational velocity.}\label{fig:mdi:vsini}
\end{figure}

We optimized the value of the projected rotational velocity in a similar manner. We carried out 13 MDI inversions with values of $v_e \sin i$ in the range from 6\,\kms{} to 19\,\kms{}. We then computed the $\chi^2$ for each value of $v_e \sin i$ for the Stokes~$I$ profiles of the Fe and Cr lines, and the \cla{rare earth elements} lines. These results are illustrated in Fig.~\ref{fig:mdi:vsini}. We found that both groups favor $v_e \sin i$ that is very close to the one derived in the abundance analysis of the average spectrum (Sect.~\ref{ssec:abund}). The minimum for the Fe and Cr lines is located at 11.4\,\kms{}, while the rare earth element lines indicate a value close to 11.1\,\kms{}. Using the solid body rotation relation and the newly derived value for the inclination, we found that $v_e \sin i = 11.4$\,\kms{} leads to $R=2.1\,R_\odot$, while the value derived from the \cla{rare earth element} lines gives $R=2.0\,R_\odot$. Both values are consistent with $R$ derived from SED fitting in Sect.~\ref{sec:fund}. Based on these results we adopted $v_e \sin i = 11.4\pm0.5$\,\kms{} for the rest of the study. 

\section{Magnetic field and chemical abundance distributions}\label{sec:mdi-maps}
\begin{figure*}
  \centering
  \includegraphics[height=\textwidth,angle=270]{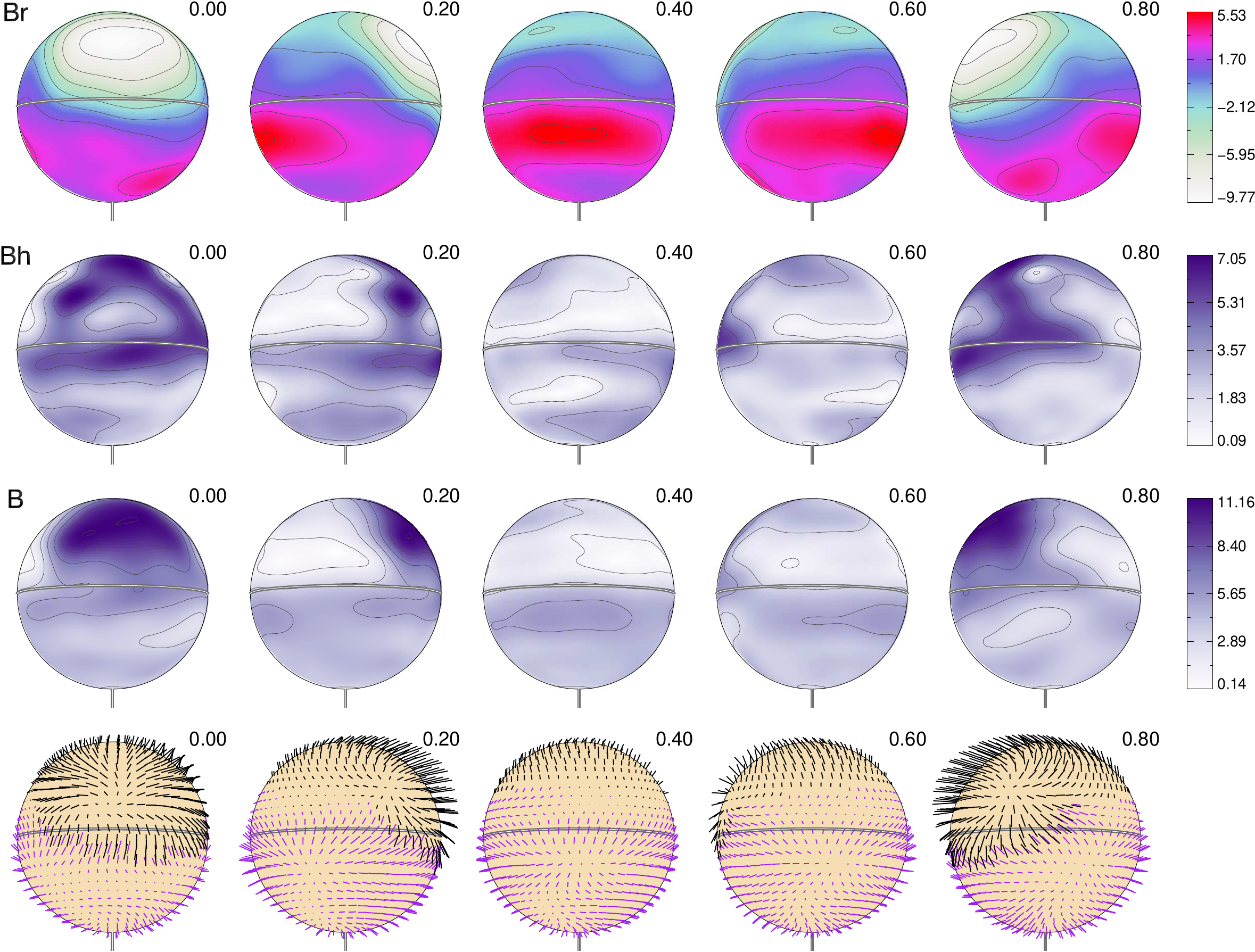}
  \caption{Distribution of the magnetic field on the surface of HD\,125248 derived from simultaneous MDI analysis of Fe, Cr, Nd, Ce, and Gd. The plots show the distribution of magnetic radial field (first row), horizontal field (second row), field modulus (third row), and field orientation (fourth row). The color bars on the right indicate the field strength in kG. The contours are plotted with 2\,kG steps. The arrow length in the bottom plot is proportional to the field strength. The star is shown at five rotational phases, indicated above each spherical plot. The thick line and the vertical bar indicate positions of the rotational equator and the pole respectively.}\label{fig:mdi:mf}
\end{figure*}

\begin{figure*}
  \centering
  \includegraphics[width=1\textwidth,page=1]{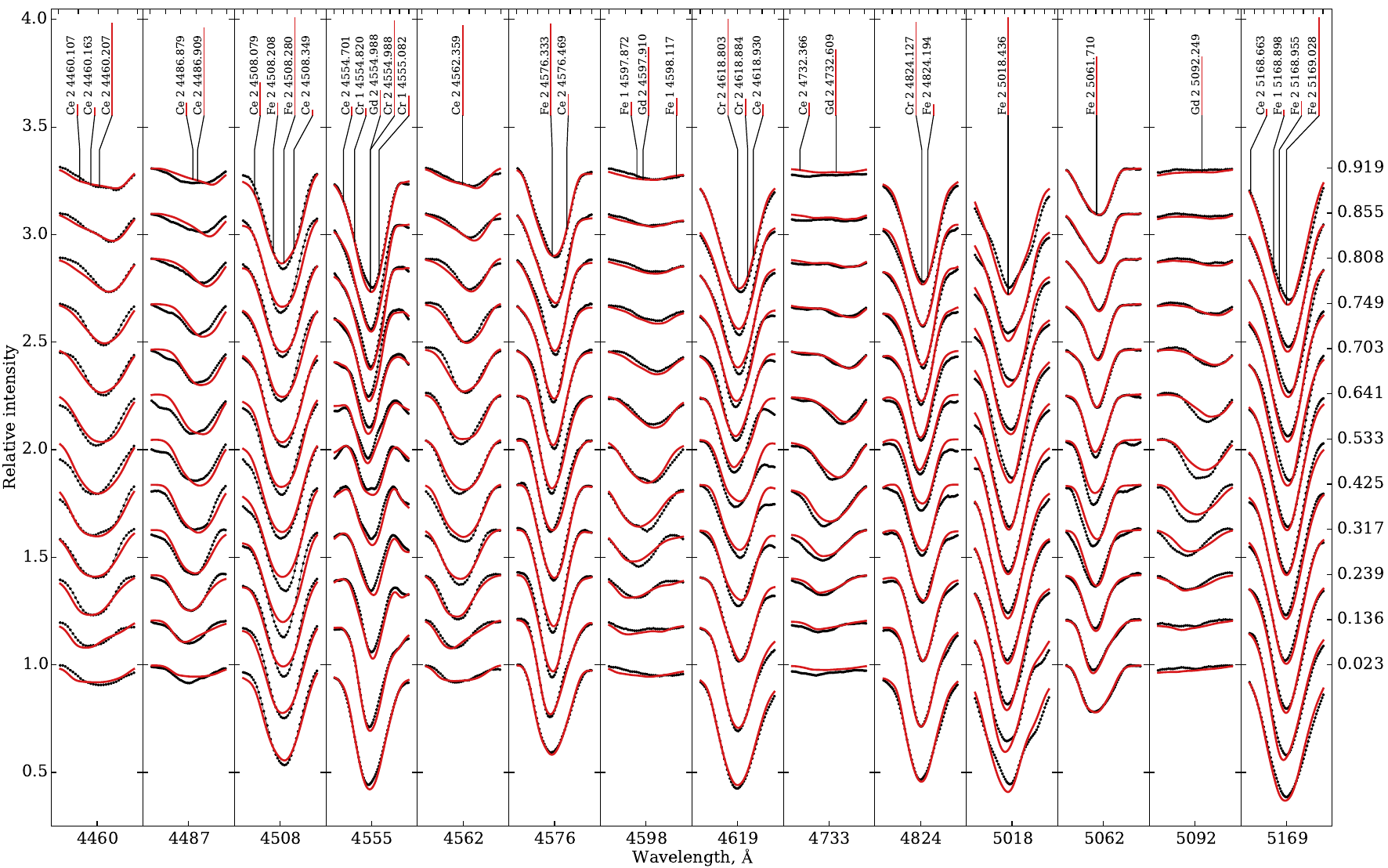}
  \includegraphics[width=1\textwidth,page=2]{{profiles_5el.DATA103}.pdf}
  \caption{Comparison of the observed (dots connected with black lines) and synthetic (thin red lines) Stokes~$I$ profiles calculated for the final magnetic field and abundance maps. The distance between two horizontal tick marks on the upper axis of each panel is 0.1\,\AA{}, indicating the wavelength scale. Rotational phases are given on the right of both panels.}\label{fig:mdi:StokesI}
\end{figure*}

The magnetic field topology of HD\,125248 was derived from the simultaneous mapping of Fe, Cr, Ce, Nd, and Gd. We fitted the phase variations of the Stokes~$IQUV$ spectra for the line list presented in Table~\ref{tab:linelist}. In Fig.~\ref{fig:mdi:mf} we plot the spherical projection of the resulting radial field component, the field modulus, and the strength of the horizontal field components; the bottom row shows the vector magnetic field. The comparison of the observed and the computed Stokes parameter spectra is presented in Figs.~{\ref{fig:mdi:StokesI}--\ref{fig:mdi:StokesV}}.

\begin{figure*}
  \centering
  \includegraphics[width=1\textwidth,page=3]{{profiles_5el.DATA103}.pdf}
  \includegraphics[width=1\textwidth,page=4]{{profiles_5el.DATA103}.pdf}
  \caption{Same as Fig.~\ref{fig:mdi:StokesI} for the Stokes~$Q$ profiles.}\label{fig:mdi:StokesQ}
\end{figure*}

\begin{figure*}
  \centering
  \includegraphics[width=1\textwidth,page=5]{{profiles_5el.DATA103}.pdf}
  \includegraphics[width=1\textwidth,page=6]{{profiles_5el.DATA103}.pdf}
  \caption{Same as Fig.~\ref{fig:mdi:StokesI} for the Stokes~$U$ profiles.}\label{fig:mdi:StokesU}
\end{figure*}

\begin{figure*}
  \centering
  \includegraphics[width=1\textwidth,page=7]{{profiles_5el.DATA103}.pdf}
  \includegraphics[width=1\textwidth,page=8]{{profiles_5el.DATA103}.pdf}
  \caption{Same as Fig.~\ref{fig:mdi:StokesI} for the Stokes~$V$ profiles.}\label{fig:mdi:StokesV}
\end{figure*}

The MDI inversions indicate that the magnetic field structure of HD\,125248 is dominated by two regions of different polarity. The area with negative polarity (inward oriented field vector) has \clb{a field modulus that is on average stronger by about 30\%} than the area with positive polarity (outward oriented field vector). We take this as an evidence of a strong asymmetry between \clb{the field strength of the positive and negative magnetic poles}. Furthermore, the area where the positive radial field reaches its maximum strength corresponds to an extended arc that almost wraps itself around the rotational pole. \clb{This can be most easily seen for the radial field component in Fig.~\ref{fig:mdi:mf} for phases 0.4 and 0.6. The area with the strongest positive radial field as traced by the contours in the plot, forms a structure that is extended in the longitudinal direction, whereas the negative radial field area is much more symmetric and localized.}

\clb{The difference in field strength between the regions of different polarity manifests itself also in their relative surface areas. This is a consequence of magnetic flux conservation which is enforced by representing the field as a superposition of a poloidal and toroidal components in the inversion. The much stronger negative field encompasses an area that is approximately two times smaller than the surface area of the field with positive polarity. }

The estimated disk-integrated mean field modulus of HD\,125248 varies between 3.5 and 6\,kG, with a large 3.5\,kG strength plateau between the rotational phases 0.3 and 0.7. This estimate of the field modulus is somewhat smaller, but still in line with previous estimates by \citet{Glagolevskij2007p244} and \citet{Bagnulo1999p865}. 

\begin{figure}
  \centering
  \includegraphics{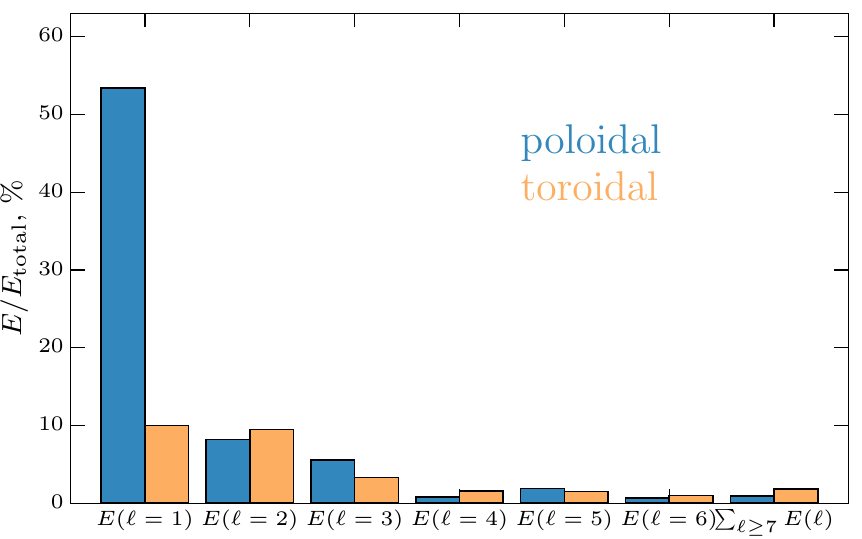}
  \caption{Relative energies of the poloidal and toroidal harmonic modes for the magnetic field topology of HD\,125248. The energy of the poloidal and toroidal components are shown in dark (blue) and light (orange) respectively. The last two bars represent the sum of energies for $\ell \geq 7$ modes.}\label{fig:mdi:energy}
\end{figure}

The analysis of the contribution of different harmonic components to the total energy of the field of HD\,125248 showed that the field is mainly poloidal (71 percent of the total magnetic energy) with a dominant contribution from the poloidal dipolar component (53 percent  of the total magnetic energy). The contribution of the toroidal components for all modes is relatively strong with 29 percent of the total magnetic energy of HD\,125248. The quadrupolar ($\ell=2$) and octupolar ($\ell=3$) components contribute 27 percent of the total energy of the field. Higher order modes with $\ell \geq 4$ contribute less than 10 percent of the total field energy. Figure~\ref{fig:mdi:energy} compares the energies of the poloidal and toroidal components as a function of the angular degree $\ell$.

\begin{figure*}
	\centering
	\includegraphics[scale=0.5, page=1]{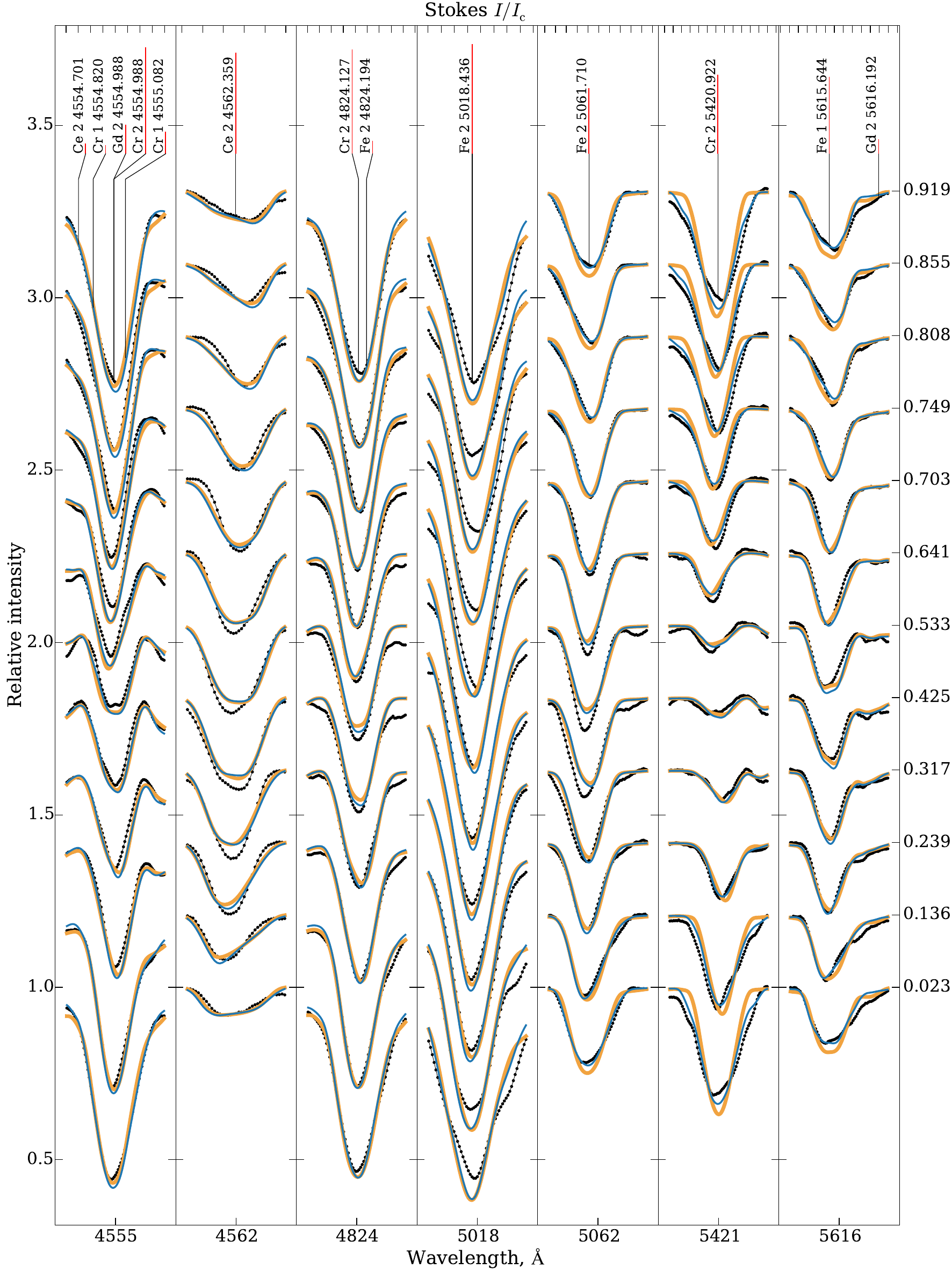}
	\includegraphics[scale=0.5, page=2]{{profiles_5el.param_models2}.pdf}\\
	\includegraphics[scale=0.5, page=3]{{profiles_5el.param_models2}.pdf}
	\includegraphics[scale=0.5, page=4]{{profiles_5el.param_models2}.pdf}
	\caption{Comparison of the observed (dots connected with black lines) and synthetic Stokes profiles calculated using dipole field (thick yellow lines), and a dipole plus quadrupole field (thin blue lines). The format of this figure is similar to Fig.~\ref{fig:mdi:StokesI}.}\label{fig:mdi:comparison}
\end{figure*}

In order to confirm that reproducing the observed Stokes profile variation indeed requires components with $\ell > 2$ we carried out inversions where we limited the maximum number of spherical harmonics for the field expansion. Our first test involved searching a best-fit dipole field ($\ell_\mathrm{max} =1$); then we fitted the observed Stokes~$IQUV$ profiles with a combination of a dipole and quadrupole field topologies ($\ell_\mathrm{max} =2$). Our best fits to the Stokes profiles are shown in Fig.~\ref{fig:mdi:comparison}. For the sake of brevity we only plotted the fit for seven spectral lines, which are representative of the entire set. Note that the inversions were done for the entire line list and the code was allowed to fit arbitrary abundance maps to the observed profiles.

It is evident that \cla{a dipole} field cannot properly reproduce the observations. In some cases it cannot even reproduce the width of certain lines for phases $\varphi \approx 0$ when the field strength is at its maximum, e.g., \ion{Cr}{ii}\,$\lambda5421$\AA{}. The theoretical Stokes~$V$ parameters are systematically underestimated for phases when the field is strongest. For several other lines the synthetic Stokes~$V$ profiles have different morphology than the observed ones. \clb{The $\chi^2$ of the fit to the Stokes~$IQUV$ profiles for the dipole field is larger by a factor of three compared to the inversion that includes higher order harmonic components.}

The dipole plus quadrupole fit on average reproduces the intensity spectra much better, keeping the theoretical Stokes~$V$ profiles close to the observed ones. Thus one can indeed say that on the largest scales the field of HD\,125248 is better represented by a combination of a dipole and quadrupole field. However, on closer inspection we can see that this model systematically fails to reproduce the Stokes~$QU$ profiles of most lines when the field is at its maximum, e.g., \ion{Fe}{ii}\,$\lambda5018$\AA{}, \ion{Cr}{ii}\,$\lambda5421$\AA{}, and \ion{Cr}{ii}\,$\lambda4824$\AA{}. This suggests that the surface field structures introduced by the harmonics with $\ell \ge 3$ are statistically significant and justified by the observational data. 

\subsection{Abundance maps}
\begin{figure*}
  \centering
  \includegraphics[width=\textwidth]{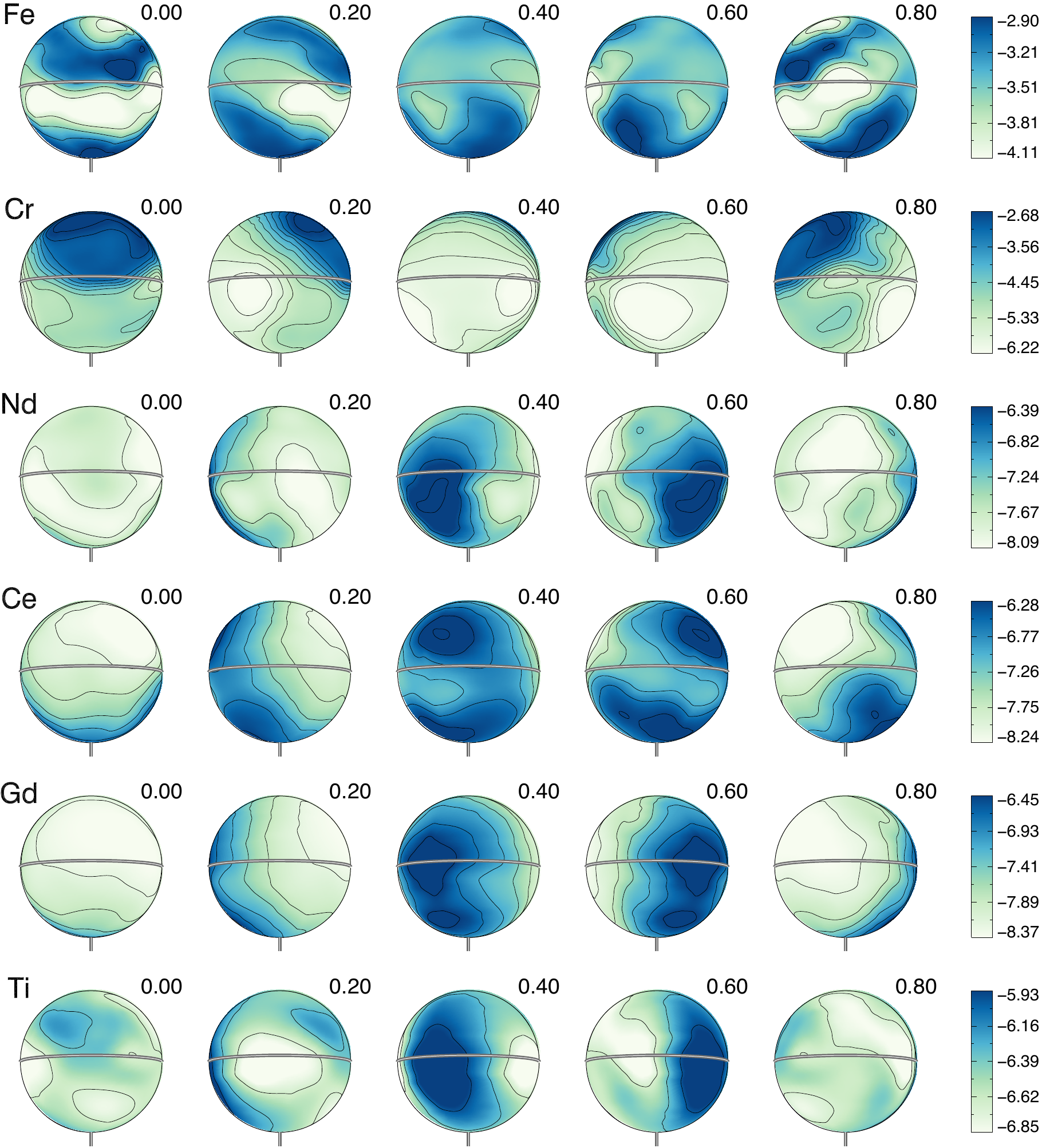}
  \caption{Abundance distributions of Fe, Cr, Nd, Ce, Gd, and Ti on the surface of HD\,125248. The color bars on the right indicate the abundance in $\log (N_X / N_\mathrm{tot})$ units. The contours are plotted with 0.35\,dex steps. The thick line and the vertical bar indicate positions of the rotational equator and the pole respectively.}\label{fig:mdi:abn}
\end{figure*}

We reconstructed the abundance distribution maps of six elements in total. The abundance maps of Fe, Cr, Nd, Ce, and Gd were obtained simultaneously with the magnetic field inversion. The abundance distribution of Ti was derived by keeping the previously inferred magnetic field geometry and abundance maps of the other chemical elements fixed. The chemical abundance maps are presented in Fig.~\ref{fig:mdi:abn}. The minimum and maximum abundance values were determined by taking the 5 and 95 percentiles for each map so that extreme outliers are excluded.

The resulting abundance maps show that HD125248 has very strong surface abundance inhomogeneities. The inferred map of Fe has abundance ranges from -2.9 to -4.1 in $\log (N_\mathrm{Fe} / N_\mathrm{tot})$ units. The abundance map shows that Fe exhibits an underabundance arc that loosely correlates with the magnetic equator. We do not see any other correlations between the magnetic field topology and the abundance map of Fe.

The chromium abundance changes most dramatically with ranges from -2.7 to -6.2 in $\log (N_\mathrm{Cr} / N_\mathrm{tot})$ units. The map of this element shows one very strong overabundance patch, which coincides with the magnetic field region of negative polarity. Such dramatic 3.5\,dex difference between the maximum and minimum value of Cr abundance is required in order to reproduce the behavior of the Stokes~$I$ profiles of lines such as \ion{Cr}{ii}\,$\lambda4555$\AA{} or \ion{Cr}{ii}\,$\lambda6134$\AA{} for phases $\varphi \approx 0.5$. Thus, it is highly unlikely that the wide range of the abundance map of Cr is an artifact of our MDI procedure, and instead it reflects what is observed in the input spectra.

The surface abundance distributions of the rare earth elements for which we performed MDI do not follow the abundance patterns of Cr or Fe. Instead \cla{rare earth elements} are concentrated where the magnetic field is of positive polarity. The abundance maps of Nd and Gd describe well the strong changes of their Stokes~$I$ profiles around phases $\varphi \approx 0$ when the line profiles of these elements become unusually shallow. The same can be said for line profiles of Ce ions, however, this element does not become strongly underabundant for phases close to zero as it is the case for Nd and Gd. For these three elements the maximum abundance appears to be close to -6.35, while their minimal value is between -8.1 and -8.4 in $\log (N_X / N_\mathrm{tot})$ units.

The abundance distribution of Ti was derived from eight \ion{Ti}{ii} lines with strong polarization signatures and noticeable rotational modulation. The selection of these lines was performed on the basis of the same principles as the main line list (Sect.~\ref{ssec:mdi:lines}). However, in this case we also included Ti lines that were slightly blended by lines of Fe, Cr, and Ce. The line selection for the mapping of Ti is presented in \cla{Table~\ref{tab:linelist-Ti}}. The abundance map of Ti was derived by using the magnetic field and abundance maps derived in the simultaneous MDI of Fe, Ce, Nd, Ce, and Gd, as fixed parameters and allowing the \invers{10} code to fit the selected Ti line profiles by finding a best-fit abundance map of this element. The resulting abundance map of Ti is illustrated in Fig.~\ref{fig:mdi:abn} (\cla{row six}). From this figure we can see that Ti does not have strong abundance contrasts; its abundance ranges from -5.9 to -6.9 in $\log (N_\mathrm{Ti} / N_\mathrm{tot})$ units. Interestingly, the surface abundance pattern of Ti follows the rare earth elements instead of Fe or Cr, a similar correlation between Ti and Nd has also been observed for $\alpha^2$\,CVn \citep{Silvester2014p1442}.

\begin{figure*}
	\centering
	\includegraphics[scale=0.5, page=1]{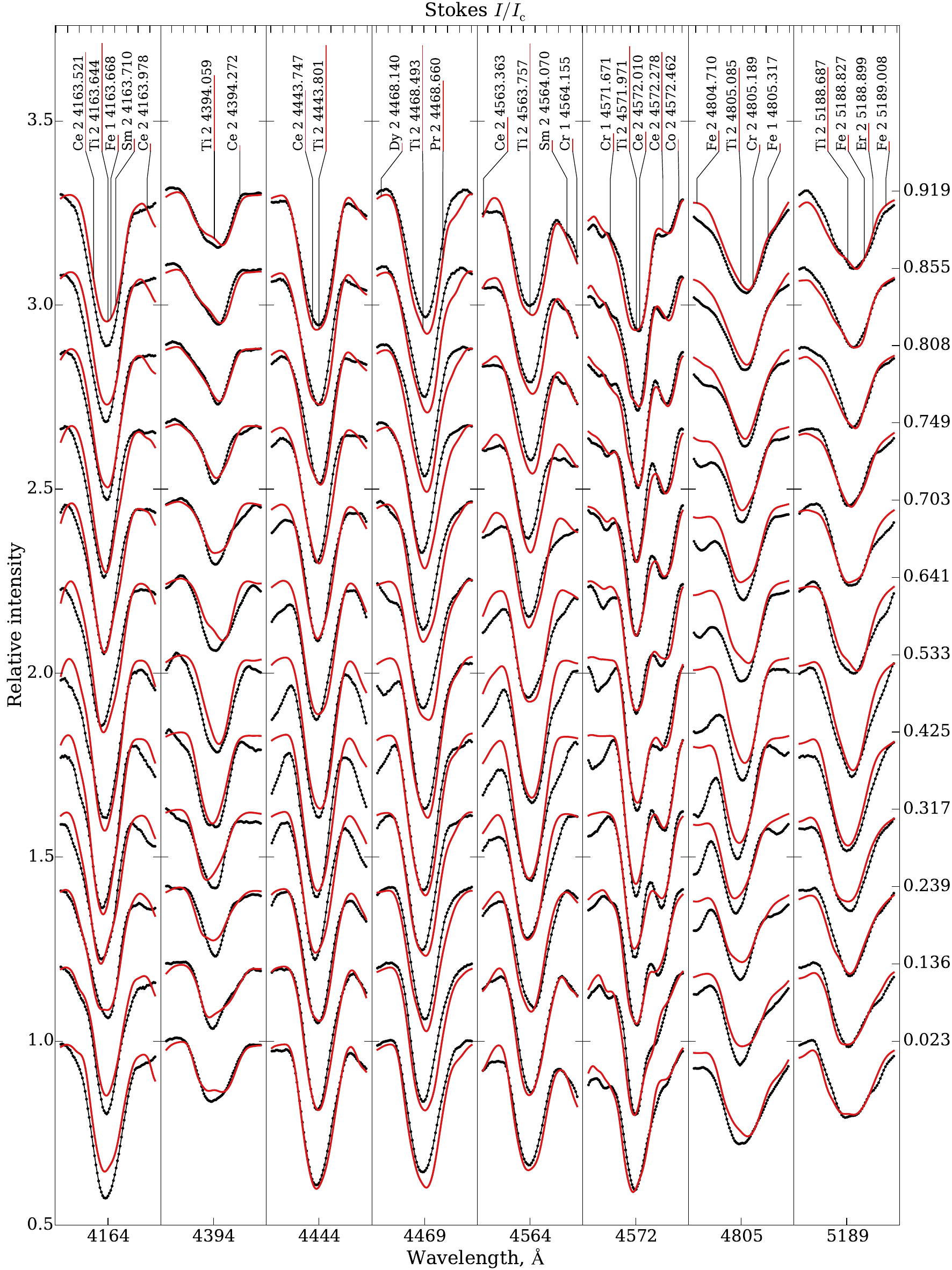}
	\includegraphics[scale=0.5, page=2]{{profiles_Ti.105}.pdf}\\
	\includegraphics[scale=0.5, page=3]{{profiles_Ti.105}.pdf}
	\includegraphics[scale=0.5, page=4]{{profiles_Ti.105}.pdf}
	\caption{Comparison of the observed (dots connected with black lines) and synthetic (thin red lines) Stokes profiles for Ti. The format of this figure is similar to Fig.~\ref{fig:mdi:StokesI}.}\label{fig:mdi:prof_Ti}
\end{figure*}

The computed and observed line profiles of Ti are shown in Fig.~\ref{fig:mdi:prof_Ti}. We can see that most polarization signatures present in the observations are reproduced by our code including the blending by spectral lines of other chemical elements. The Stokes~$Q$ profiles of several Ti lines for phase $\varphi \approx 0.136$ are the only example of a systematic discrepancy that we could find. This happens close to the phase when we see a sign change of the Stokes~$U$ profiles for several lines, e.g., \ion{Fe}{ii}\,$\lambda5018$\AA{} and \ion{Fe}{ii}\,$\lambda5169$\AA{}. Similar behavior around this phase can also be noticed for the Ti Stokes~$QU$ profiles. We do not consider these minor discrepancies capable of invalidating our results for the following reasons. There are no strong systematic discrepancies between the observed and synthetic line profiles from which we determined the magnetic field structure. The synthetic Stokes~$Q$ profiles for Ti have weaker amplitudes at phase 0.136, while their morphology is correctly reproduced for other phases.

\begin{table}
\caption{Atomic data used in the abundance mapping of titanium.}\label{tab:linelist-Ti} 
\centering
\begin{tabular}{lccccc}
\hline
\hline
Ion & $\lambda$\,(\AA{})& $E_\mathrm{i}$\,(eV)& $\log gf$& $\bar{g}$\\
\hline
\ion{Ti}{ii}& 4163.644&  2.590&  -0.13& 1.07\\ 
\ion{Ti}{ii}& 4394.059&  1.221&  -1.77& 1.34\\ 
\ion{Ti}{ii}& 4443.801&  1.080&  -0.71& 0.92\\ 
\ion{Ti}{ii}& 4468.493&  1.131&  -0.63& 1.05\\ 
\ion{Ti}{ii}& 4563.757&  1.221&  -0.69& 0.99\\ 
\ion{Ti}{ii}& 4571.971&  1.572&  -0.31& 0.94\\ 
\ion{Ti}{ii}& 4805.085&  2.061&  -0.96& 1.14\\ 
\ion{Ti}{ii}& 5188.687&  1.582&  -1.05& 1.20\\ 
\hline
\end{tabular}
\tablefoot{Same as Table~\ref{tab:linelist}.}
\end{table}

\section{Summary and discussion}\label{sec:sum}
We have obtained spectra in all four Stokes parameters of the magnetic Ap star HD\,125248. The spectropolarimetric observations were obtained with the HARPSpol instrument at the \cla{3.6-m} ESO telescope, and have resolving power exceeding $10^5$ and S/N ratio of 200--300. The full set of spectropolarimetric observations is comprised of 36 individual Stokes parameter spectra and covers the entire rotational period of the star.

Based on these HARPSpol spectra we calculated LSD profiles using different line masks. These profiles were used to obtain new precise measurements of the mean longitudinal magnetic field and net linear polarization. The longitudinal field measurements were combined with literature data to find an improved rotational period. 

The atmospheric parameters of HD\,125248 were derived with the help of modern atmosphere models that include detailed treatment of non-solar abundances and the presence of a strong magnetic field. Our abundance analysis of a large number of chemical elements showed that HD\,125248 is deficient in He and shows overabundances for most Fe-peak and rare earth elements. The abundance analysis of  \ion{Fe}{ii} and \ion{Cr}{ii} lines indicates that HD\,125248 has vertical abundance gradients in its atmosphere. We did not find strong difference in the abundances measured for different ions of the \cla{rare earth elements} reported for cooler Ap stars by \citet{Ryabchikova2004p705}.

The magnetic Doppler imaging of HD\,125248 revealed a magnetic field that is dipole-like only on the largest spatial scales, with strong field strength asymmetry between the two regions of opposite polarity and a major deviation from axisymmetry. Most of the magnetic field energy of HD\,125248 is contained in the $\ell$~=~{1--3} harmonic modes. Our analysis showed that neither \cla{a purely dipole field} nor a more complex dipole plus quadrupole field topology can reproduce the observations in all Stokes parameters. 

The abundance distribution maps derived in the MDI analysis showed high contrast regions. However, we do not see obvious correlation between horizontal field strength and overabundance patches as suggested by the recent theoretical studies of atomic diffusion in the presence of magnetic fields \citep{Alecian2010p53,Alecian2015p3143}. Instead we see that Cr is most abundant where the field is of negative polarity, while the rare earth elements follow the field with positive polarity. The abundance of Fe does not correlate with the magnetic field in any obvious way except for a relative underabundance in the area of horizontal field.

\citet{Silvester2015p2163} made a qualitative assessment of the complexity of the magnetic field of several Ap stars by comparing the distribution of the magnetic field energy derived from MDI studies as a function of angular degree to other stellar parameters. They found that a correlation between field complexity and stellar mass or age is probably present, as already suggested by \citet{Rusomarov2015p123}.

Our results indicate that the magnetic field complexity of HD\,125248 is similar to CU\,Vir \citep{Kochukhov2014p83}, $\alpha^2$\,CVn \citep{Silvester2014p182}, and HD\,32633 \citep{Silvester2015p2163}. These four stars are young objects with masses between 2.3 and 3.1\,$M_\odot$. \citet{Rusomarov2015p123} found \cla{a dipole field} for HD\,24712, which has a mass of 1.55\,$M_\odot$ and is likely to be much older than the other stars. Thus, it appears that our finding of magnetic field for HD\,125248 that is significantly more complex than a dipole supports the hypothesis that old (or less massive) Ap stars have predominantly dipolar fields with little structure on small scales. However, given the small sample of Ap stars for which full Stokes vector MDI studies have been carried out, it is difficult to make any definite conclusions about these trends. Nonetheless, as more four Stokes parameter MDI studies become available, we will be able to fully assess correlations between the field complexity and stellar parameters of magnetic Ap stars.

\begin{acknowledgements}
OK acknowledges financial support from the Knut and Alice Wallenber Foundation, the Swedish Research Council, and the G\"oran Gustafsson Foundation. The computations presented in this paper were performed on resources provided by SNIC through Uppsala Multidisciplinary Center for Advanced Computational Science (UPPMAX) under project p2013234. TR acknowledges partial financial support from the Presidium of RAS Program P-41. Resources provided by the electronic databases (VALD, Simbad, NASA ADS) are gratefully acknowledged. NR is profoundly grateful to J.~Silvester for the deeply stimulating discussions. 
\end{acknowledgements}


\end{document}